\documentclass[sigconf,10pt]{acmart}
\AtBeginDocument{%
  }

\setcopyright{acmlicensed}
\copyrightyear{2018}
\acmYear{2018}
\acmDOI{XXXXXXX.XXXXXXX}
\acmConference[Conference acronym 'XX]{}{}{}
\acmISBN{978-1-4503-XXXX-X/2018/06}

\usepackage[utf8]{inputenc}

\begin{document}
\newcommand{\placeholder}{MelodyBedScale }
\title{Human Body Weight Estimation Through Music-Induced Bed Vibrations}

\author{Yuyan Wu}
\email{wuyuyan@stanford.edu}
\orcid{0009-0009-3152-939X}
\affiliation{%
  \institution{Stanford University}
  \city{Stanford}
  \state{California}
  \country{USA}
}

\author{Jiale Zhang}
\affiliation{%
  \institution{Dept of Electrical Engineering and Computer Science, University of Michigan}
  \city{Ann Arbor}
  \state{Michigan}
  \country{USA}}

\author{Moon Lee}
\affiliation{%
  \institution{Dept of Emergency Medicine, Stanford University}
  \city{Stanford}
  \state{California}
  \country{USA}
}

\author{Cherrelle Smith}
\affiliation{%
  \institution{Dept of Emergency Medicine, Stanford University}
  \city{Stanford}
  \state{California}
  \country{USA}
} 

\author{Xinyi Li}
\affiliation{%
  \institution{Dept of Civil Engineering, Stanford University}
  \city{Stanford}
  \state{California}
  \country{USA}
}

\author{Ankur Senapati}
\affiliation{%
  \institution{Purdue University}
  \city{West Lafayette}
  \state{Indiana}
  \country{USA}
}

\author{Pei Zhang}
\affiliation{%
  \institution{Dept of Electrical Engineering and Computer Science, University of Michigan}
  \city{Ann Arbor}
  \state{Michigan}
  \country{USA}}
  
\author{Hae Young Noh}
\affiliation{%
  \institution{Dept of Civil Engineering, Stanford University}
  \city{Stanford}
  \state{California}
  \country{USA}
}
\renewcommand{\shortauthors}{XXX et al.}

\begin{abstract}
Rapid and accurate body weight estimation is critical in emergency medical care, as it directly influences treatment decisions, such as drug dosing, defibrillation energy selection, and fluid resuscitation. Traditional methods such as stand-on scales, length-based tapes, or transfer-based weighing scales are often impractical for immobilized patients, inaccurate, or labor-intensive and time-consuming. This paper introduces MelodyBedScale, a non-intrusive and rapid on-bed weight estimation system that leverages bed vibration induced by music. The core insight is that body weight affects the vibration transfer function of the bed-body system, which is captured using vibration sensors placed on opposite sides of the bed. First, we identify weight-sensitive frequency bands and compose clinically acceptable soft, natural music with high signal energy in these frequency bands. This music is then played through a speaker mounted on the bed to induce bed vibrations. Additionally, to efficiently capture the complex weight–vibration relationship with limited data and enhance generalizability to unseen individuals and weights, we theoretically analyze the weight–vibration relationship and integrate the results into the neural network’s activation functions for physics-informed weight regression. 
We evaluated MelodyBedScale on both wooden and steel beds across 11 participants, achieving a mean absolute error of up to 1.55 kg (97.6$\%$ accuracy).
\end{abstract}




\maketitle
\section{Introduction}
Accurate and timely body weight information is essential in emergency medicine for calculating drug dosages (e.g., sedatives, anesthetics)\cite{krauss2006procedural}, determining fluid resuscitation volumes in burn patients\cite{massEMS2025}, setting electrical energy levels for cardioversion or defibrillation~\cite{vinter2023electrical}, and guiding fluid replacement in trauma or dehydration cases~\cite{castera2025fluid}. Inaccurate weight measurements can result in serious clinical consequences, including organ failure, seizures, or treatment failure~\cite{authority2009medication, greenwalt2017elimination}. Clinical safety standards require that 70$\%$ of estimates fall within ±10$\%$ of actual weight and 95$\%$ within ±20$\%$~\cite{wells2017accuracy}. However, obtaining accurate and rapid weight measurements is challenging in emergency settings, as severely injured or unconscious patients often cannot use conventional weighing scales due to mobility limitations.

Current body weight estimation methods for immobilized patients in emergency scenarios include visual assessment, length-based tapes (e.g., Broselow Tape), transfer-based weighing scales, and on-bed load cell systems. Visual assessment relies on medical professionals' subjective evaluation of patient body size, yielding highly inaccurate estimation that typically deviates by 10–20 kg from actual weight, which cannot meet clinical standards~\cite{anglemyer2004accuracy}. Length-based tapes estimate weight from body height measurements~\cite{lubitz1988rapid,wells2013pawper,abdel2012improved}. However, they fail to account for individual body shape variation, making it inaccurate for patients who are underweight or overweight~\cite{wells2017accuracy, wells2023there, mehta2020accuracy}. Transfer-based methods, including under-bed scales and pressure mats~\cite{healthometer2019accuracy, charder2024ms6001, wu2023massnet}, require the movement of the patient or bed, which may increase safety risks, require additional staff, and delay urgent care~\cite{vieira2009safety}. While on-bed load cell systems provide accurate, transfer-free, and accurate body weight measurements, they are usually costly, require patient displacement for calibration, and are difficult to retrofit~\cite{shafi2022design}. These limitations highlight the critical need for rapid, non-intrusive, and accurate weight estimation methods that can be deployed without patient displacement in emergency care settings.

\begin{figure}[t]
    \centering
    \includegraphics[width=0.9\linewidth]{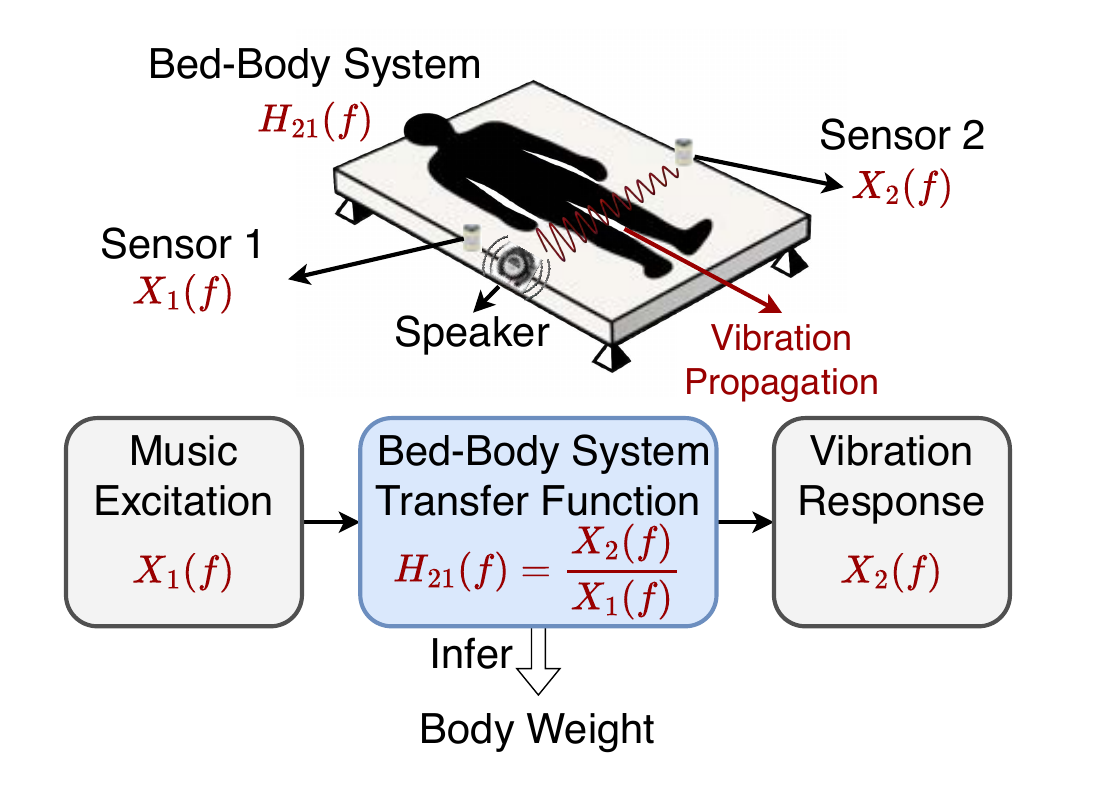} 
    \caption{Main intuition of body weight estimation through music-induced bed vibrations. The body weight is inferred from the vibration transfer function of the bed-body system, which is captured using vibration sensors placed on opposite sides of the bed.}
    \label{fig:intuition}
\end{figure}

This paper introduces MelodyBedScale, a novel system for estimating human body weight through music-induced bed vibrations. The main intuition of MelodyBedScale is that body weight affects the vibration transfer function of the bed–body system by changing its effective mass (see Figure~\ref{fig:intuition}). To capture this transfer function change, we actively induce the bed vibrations with music played through a speaker attached to the bed. The music-induced bed vibrations propagate through the bed-body system and are then captured by vibration sensors deployed on the opposite side of the bed. By computing the transfer function with the music excitation signal and the measured vibration response, human body weight can be inferred.

However, achieving this goal presents four main research challenges. First, the relationship between body weight and bed vibrations is complex and indirect, confounded by bed structural properties such as bed dimensions, boundary conditions, and material properties that are difficult to measure and quantify accurately and directly. Second, collecting data across the full range of body weights is costly, which often results in limited datasets with sparse coverage of body weight values. The model needs to generalize effectively to unseen weight values and individuals despite the limited training dataset. Third, acoustic excitation must comply with clinical safety regulations while inducing distinctive vibration responses that vary with weight, which restricts the feasible sound options. 
Finally, variations in the bed–body contact area influence vibration responses along with body weight. Because this contact area differs across individuals, its effect is difficult to capture, making it challenging to disentangle its contribution from that of body weight.

To address these challenges, we develop the MelodyBedScale system with corresponding solutions. First, we theoretically formulate and characterize how human body weight affects the vibration transfer function spectrum of the bed-body system through structural dynamics analysis. This analysis provides the functional form of the weight–vibration relationship and reveals the confounding effect of unknown structural properties. We incorporate these insights into a physics-informed neural network with specialized activation functions derived from the theoretical results, representing the unknown structural properties as learnable parameters within the network. This integration of physical principles with the neural network enables a data-efficient and generalizable model for estimating body weight. Additionally, we identify weight-sensitive frequency bands for each type of bed by inducing bed vibrations with a chirp signal and extracting the high-energy response frequencies.
We then compose natural, soft music with high energy in the identified weight-sensitive frequency bands using an AI-based music composition tool and employ this music to induce bed vibrations for weight estimation.
Finally, we incorporate body height as an indicator of the bed–body contact area into the regression model through a theory-guided activation function, enabling the disentanglement of contact-area effects as a confounding factor.

We evaluated MelodyBedScale on wooden and steel beds with 11 participants each. Bed vibrations were induced by an audio speaker and captured using two vibration sensors per deck board to compute transfer functions. Using leave-one-person-out cross-validation, MelodyBedScale achieved a mean absolute error (MAE) of 1.55 kg (2.37$\%$) on the wooden bed and 4.36 kg (6.38$\%$) on the steel bed. 
Both results satisfy clinical acceptability standards. 

The contributions of this paper are summarized as follows:

\begin{enumerate}
\item  We develop a novel human body weight estimation system, MelodyBedScale, through music-induced bed vibrations. Our system provides accurate, rapid, and non-invasive weight measurements.

\item  We formulate the complex and indirect relationship between human body weight and bed vibration response through structural dynamics theory and develop a physics-informed machine learning approach for data-efficient and interpretable weight estimation.

\item  We characterize the weight-sensitive frequency band and compose effective music excitations to induce weight-sensitive bed vibrations while satisfying clinical requirements.

\item  We evaluated MelodyBedScale through real-world experiments on two types of beds, each with 11 participants, achieving performance with estimation errors as low as 1.55 kg (2.37$\%$), satisfying clinically acceptable standards.
\end{enumerate}

The remainder of this paper is organized as follows: Section~\ref{sec:theory} presents the physical insight and theoretical analysis of the relationship between human body weight and bed vibrations, which informs the physics‑informed regression model in MelodyBedScale. Section~\ref{sec:method} details the MelodyBedScale system. Section~\ref{sec:evaluation} describes the human experiments conducted on two bed types and reports the results. 
Section~\ref{sec:related_works} reviews related work. Finally, Section~\ref{sec:conclusion} concludes the paper and outlines potential future research directions.

\section{Physical Insight and Theoretical Analysis of Weight–Vibration Relationship}
\label{sec:theory}
This section introduces the main intuition of MelodyBedScale, followed by a theoretical analysis of the weight–vibration relationship, which inspires the development of the physics-informed body weight regression model.

The main intuition of our system is that the human body weight changes the vibration transfer function of the bed-body system by introducing additional mass. These changes can be captured by vibration sensors placed on the opposite sides of the bed. An illustrative analogy is the drum: when the drum is struck in its unloaded state, the drum produces a crisp and resonant sound. But if we place a heavy object on the drumhead and strike it again, the sound becomes deeper or more muted. This is because the added mass affects the dynamic response of the drum–object system. The added object affects multiple vibrational characteristics, including frequency content, amplitude, resonance behavior, and damping properties. Similarly, human body weight affects the vibration transfer function of the bed–body system, which enables us to infer body weight through the analysis of the bed vibration responses.

We formulate the relationship between human body weight and bed vibration response through structural dynamics analysis. The bed is modeled as a rectangular plate, and the human body is represented as an added static mass $m_0$. Music excitation is modeled as a point force with time history \( f_e(t) \), applied at \( (x_e, y_e) \). The frequency-domain representation of this excitation is denoted as $X_1(\omega) = \hat{f}_e(\omega)$, which is the signal input to the bed-body system. The vertical bed velocity response, measured by the vibration sensor, is obtained from the governing equation (Equation~\ref{eq:vibration})~\cite{humar2012dynamics, fryba2013vibration}, which describes a plate vibrating subjected to the static load $m_0$ and the excitation force \( f_e(t) \) from the speaker. The plate’s bending rigidity is given by \( D = \frac{Eh^3}{12(1-\nu^2)} \), where \( h \) is the thickness, \( E \) is Young's modulus, and \( \nu \) is Poisson's ratio. The mass per unit area is denoted by \( \mu \). Vertical displacement at position \( (x, y) \) and time \( t \) is represented by \( w(x,y,t) \), with velocity vibration response \(v(x,y,t) = \dot{w}(x,y,t)\). Damping is characterized by the circular frequency \( \omega_b \). The bed-body contact area is defined by \( U(S) \) (equal to 1 within the contact region and 0 elsewhere). \( S \) represents the area size.

\begin{align}
    &D\nabla^4 w(x, y, t) + \mu\frac{\partial^2 w(x, y, t)}{\partial t^2} 
    + 2\mu \omega_b \frac{\partial w(x, y, t)}{\partial t} \notag \\
    &= f_e(t)\delta(x-x_e)\delta(y-y_e) 
    + \frac{U(S)(m_0g - m_0\frac{\partial^2w(x,y,t)}{\partial t^2})}{S} 
\label{eq:vibration}
\end{align}

The relationship between the frequency-domain vibration response and the human body weight $m_0$ can be represented using a Padé approximant. Solving the governing equation~\ref{eq:vibration}, the frequency-domain velocity vibration response at a sensor location $(x,y)$ is written as
\begin{align}
    v(x,y,\omega) &= i\omega w(x,y,\omega) = i\omega\sum_{m=1}^{\infty}\sum_{n=1}^{\infty}\phi_{mn}(x,y)\hat{q}_{mn}(\omega) \notag \\
    &= \sum_{m=1}^{\infty}\sum_{n=1}^{\infty}
    \frac{i\omega\hat{f}_e(\omega)\phi_{mn}(x_e,y_e)\phi_{mn}(x,y)}
    {-\omega^2(\mu C_2+\frac{m_0}{S}C_3)+2i\omega\mu\omega_b C_2+DC_1}.
    \label{eq:vibration_resp}
\end{align}

The amplitude of the frequency-domain velocity vibration response, referred to as $X_2(\omega) =|v(x,y,\omega)|$, as shown in Equation~\ref{eq:vibration_resp}, varies with the object mass $m_0$ as follows:
\begin{equation}
    X_2(\omega) = |\sum_{m=1}^{\infty}\sum_{n=1}^{\infty} \frac{X_1(\omega)A_{mn}(\omega)}{m_0 + B_{mn}(\omega)}|\approx X_1(\omega)\frac{\sum_{i=0}^{4} a_i m_0^i  }{1+\sum_{j=1}^{4} b_j m_0^j}
    \label{eq:freq_solu}
\end{equation}
where
$A_{mn}(\omega) = \frac{\phi_{mn}(x_e,y_e)\phi_{mn}(x,y)S}{-\omega C_3},$
$B_{mn}(\omega) = -\frac{S}{\omega^2 C_3}(DC_1 - \omega^2 \mu C_2 + 2i\omega \mu \omega_b C_2)$.
Here, indices $m$ and $n$ denote structural vibration mode numbers. Although the theoretical solution involves summation over infinite modes, in practice, vibration response is predominantly governed by the first few modes (typically fewer than five). Higher-order modes contribute negligibly. Consequently, the infinite series can be effectively approximated by a rational polynomial form, known as the Padé approximant, as shown in Equation~\ref{eq:freq_solu}.

The bed-body contact area influences the vibration response through the integral of the squared mode shape function $\phi_{mn}(x,y)$ over this area. This effect is encapsulated in the coefficient $C_3$, given by: $C_3 = \int \int_{S}\phi_{mn}^2(x,y)dxdy$. The mode shape function $\phi_{mn}(x,y)$ typically consists of trigonometric and hyperbolic functions and generally can be decomposed into the product form: $\phi_{mn}(x,y) = \phi_{m}(x)\phi_n(y)$, with each component expressed as: $\phi_{\xi} (\xi) = A_\xi sin(\nu_{0\xi} \xi)+ B_xcos(\nu_{1\xi} \xi)+C_\xi sinh(\nu_{3\xi}\xi)+D_\xi cosh(\nu_{4\xi}\xi), \xi \in \{x, y\}, \phi_{\xi}\in \{\phi_m, \phi_n\}$. Under simply supported boundary conditions, this expression simplifies to: $\phi_\xi (\xi) = sin(\nu_\xi \xi)$. This sinusoidal form motivates the selection of learnable sine activation functions to incorporate height information (an indicator of bed-body contact area) within the physics-informed weight regression model.

The weight-sensitive frequency bands are located near the natural frequencies of the bed structure. Weight sensitivity is quantified by the derivative of the transfer function $H(\omega)$ (Equation~\ref{eq:transfer}) with respect to body weight $m_0$, as shown in Equation~\ref{eq:sensitivity}. The natural frequencies correspond to the values of $\omega$ that satisfy $m_0 + B_{mn}(\omega) = 0$, i.e., the zeros of the transfer function $H(\omega)$. Near these frequencies, the denominator of $\frac{dH(\omega)}{dm_0}$ is also small, indicating high sensitivity. Consequently, the high sensitivity occurs at the bed’s natural frequencies. This observation enables the identification of effective music capable of inducing weight-sensitive vibrations by capturing the natural frequency bands.

\begin{equation}
    H(\omega) = \frac{X_2(\omega)}{X_1(\omega)}=|\sum_{m=1}^{\infty}\sum_{n=1}^{\infty} \frac{A_{mn}(\omega)}{m_0 + B_{mn}(\omega)}|\approx\frac{\sum_{i=0}^{4} a_i m_0^i  }{1+\sum_{j=1}^{4} b_j m_0^j}
    \label{eq:transfer}
\end{equation}
\begin{equation}
    \frac{dH(\omega)}{dm_0} = |\sum_{m=1}^{\infty}\sum_{n=1}^{\infty}\frac{A_{mn}(\omega)}{(m_0 + B_{mn}(\omega))^2}|
    \label{eq:sensitivity}
\end{equation}

These theoretical results inform the development of physics-informed activation functions within the body weight regression model. Additionally, the results enable identification of weight-sensitive frequency bands by characterizing the natural frequency bands of the bed structure. These weight-sensitive frequency bands guide both music excitation composition and feature selection for transfer function spectra in the MelodyBedScale system.



\section{MelodyBedScale System}
\label{sec:method}
\begin{figure*}[t]
    \centering
    \includegraphics[width=\textwidth]{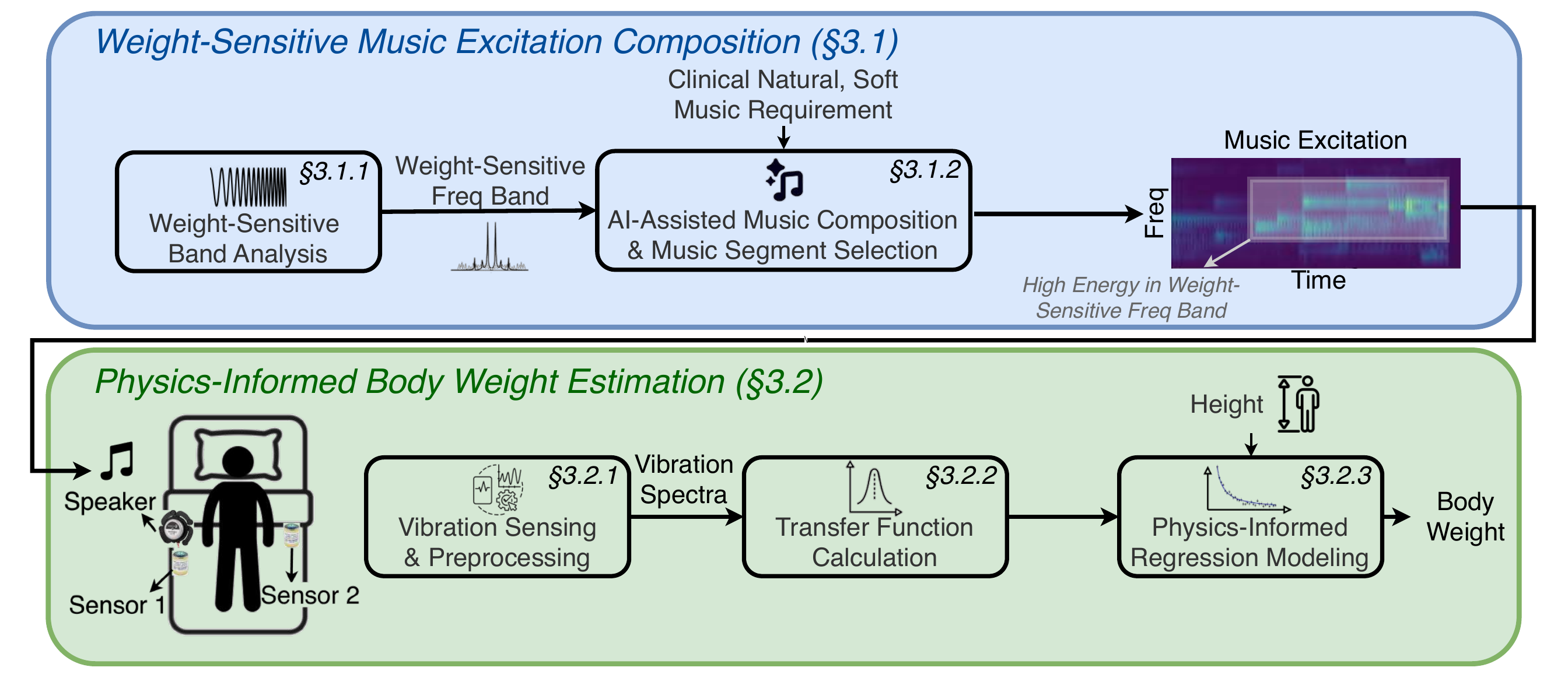}
    \caption{System overview of MelodyBedScale.}
    \label{fig:system}
\end{figure*}

The MelodyBedScale system includes two main modules: Weight-Sensitive Music Excitation Composition (Section~\ref{sec:music_gen}) and Physics-Informed Body Weight Estimation (Section~\ref{sec:phy_model}). The system overview is shown in Figure~\ref{fig:system}. In the first module, we identify the weight-sensitive frequency band for the bed structure and compose music that includes high energy within these frequency bands to induce bed structural vibrations, assisted with an AI song-composition tool. In the second module, the bed structural vibrations are captured and analyzed to develop a physics-informed regression model for body weight estimation.

\subsection{Weight-Sensitive Music Excitation Composition}
\label{sec:music_gen}
The choice of music excitation must induce effective bed structural vibrations that are differentiable for different weights while ensuring compliance with hospital acoustic regulations. The chirp signal, which spans a broad frequency range, is ideal for inducing weight-sensitive bed vibrations. However, its alarm-like acoustic characteristics may cause patient discomfort and are unsuitable for clinical environments. Instead, only soft, natural music that is non-disruptive and comforting to patients is allowed. To address this, we first identify the weight-sensitive frequency bands of the bed structure by analyzing its natural frequency bands through chirp-induced resonant vibrations. This identification is performed once for each bed-structure type and is not required during actual weight estimation for each patient. The weight-sensitive frequency band information, combined with hospital acoustic requirements, is used to formulate a prompt for AI-assisted music composition. Finally, we extract suitable segments from the AI-composed music based on its energy within the weight-sensitive frequency bands.

\subsubsection{Weight-Sensitive Frequency Band Analysis}
\label{sec:freq_band_sele}

We identify the weight-sensitive frequency band by characterizing the natural frequency band of the bed using chirp excitation. As proven in Section~\ref{sec:theory}, the weight-sensitive frequency band corresponds to the natural frequencies of the bed structure. To determine this band, we induce the bed vibrations with a chirp signal of uniform amplitude spanning 10–1000 Hz over 1 second and record the chirp-induced bed vibrations. The measured vibrations exhibit higher energy within the natural frequency band due to resonant amplification, thereby directly identifying the weight-sensitive frequency band.


\subsubsection{AI-Assisted Music Composition and Music Segment Selection}

To obtain music excitation which satisfies clinical requirements for naturalness and softness while maintaining high energy within weight-sensitive frequency bands, we employ the AI-based composition tool SunoAI~\cite{sunoai2023}. The weight-sensitive frequency bands are first mapped to Scientific Pitch Notation, after which style constraints for softness and naturalness are incorporated into the prompt along with the pitch notation. This prompt is provided to SunoAI to generate a music clip. Because SunoAI produces three-minute outputs, we extract the most effective 10-second segment identified by the highest energy within the weight-sensitive frequency band. This segment is then used as the music excitation to induce bed vibrations.

\subsection{Physics-Informed Body Weight Estimation}
\label{sec:phy_model}
The physics-informed body weight estimation framework comprises three steps: vibration sensing and preprocessing, vibration transfer function calculation, and physics-informed regression modeling. As shown in Section~\ref{sec:theory}, vibration transfer function spectra are effective features for body weight regression, as they vary with different body weights applied on the bed. The transfer function of each deck board is calculated with music-induced vibrations measured on opposite sides of the bed. The concatenated transfer functions from all deck boards, along with the height information indicating the bed-body contact area, serve as input to a physics-informed regression neural network. This network employs physics-guided activation functions that parameterize unknown structural property confounders as learnable variables, enabling data-efficient and interpretable body weight estimation.

\subsubsection{Vibration Sensing and Preprocessing}
MelodyBedScale captures the bed vibrations with synchronized geophone vibration sensors mounted on the deck boards of the bed. The transfer function of each deck board is obtained from two sensors: one positioned on the side with the speaker to measure the excitation input to the bed–body system and the other on the opposite side to record the vibration response after propagation through the system. Both sensors measure the vertical velocity vibrations of the deck boards. Both the speaker and sensors are installed at the side of the bed, ensuring they do not interfere with normal bed usage. For better sensitivity, the sensors are positioned as far as possible from support points (the connections between bed feet or wheels and deck boards). This is because vibration amplitudes are attenuated at these points due to the direct force transmission to the ground, resulting in a reduced sensor signal-to-noise ratio.


\subsubsection{Vibration Transfer Function Calculation}
The vibration transfer function for each deck board is calculated from frequency spectra obtained by two vibration sensors positioned on opposite sides of the board. The transfer function quantifies the bed system's vibration frequency response and is defined as the ratio of system output to input. In the bed-body system, the input corresponds to the music excitation, while the output corresponds to the vibration response on the side opposite to the speaker. The vibration sensor on the speaker side, which captures the input signal, is referred to as sensor 1, while the sensor on the opposite side is referred to as sensor 2. The power spectral density (PSD) of sensor 1 is estimated using Welch's method and denoted as $P_{xx}(f)$, where $f$ represents the frequency. The cross-spectral density between sensor 2 and sensor 1 is computed as $P_{xy}(f)$. The transfer function is then given by $H_{21}(f) = \frac{P_{xy}(f)}{P_{xx}(f)}$. This approach provides more robust estimates than direct spectral division by mitigating random noise effects (e.g., electrical interference, ambient vibrations, and sensor imperfections)~\cite{bendat1986random}. The resulting transfer function spectrum is band-pass filtered to reserve only weight-sensitive frequencies and used as the feature input for the physics-informed body weight regression model.


\subsubsection{Physics-Informed Body Weight Regression Modeling}
\label{sec:phy_method}

\begin{figure}[t]
    \centering
    \includegraphics[width=0.5\textwidth]{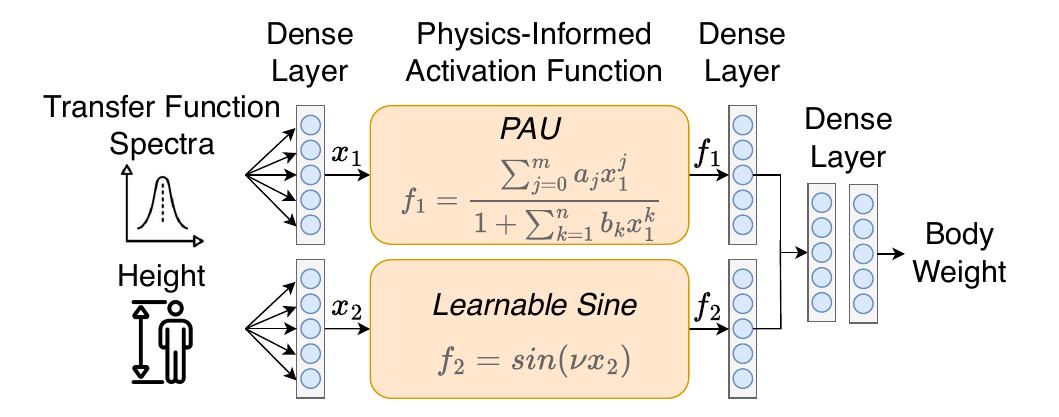} 
    \caption{Architecture of the physics-informed neural network for body weight estimation.}
    \label{fig:network}
\end{figure}

The body weight regression model incorporates theoretical insights into the network architecture, taking the vibration transfer function spectra and body height as inputs and providing the estimated body weight as output.
The physics-informed body weight regression model is intentionally simple yet efficient, consisting solely of dense layers with physics-informed activation functions (see Figure~\ref{fig:network}).
The model inputs are the concatenated vibration transfer function spectra of all deck boards and the measured human height. Its shallow architecture is designed to mitigate overfitting, as the number of distinct weight labels is constrained by the limited number of individuals in the dataset.


Inspired by the theoretical analysis in Section~\ref{sec:theory}, which indicates that the relationship between human body weight and the vibration transfer function spectra follows the form of a Padé approximant, we adopt a physics-informed activation function, the Padé Activation Unit (PAU)~\cite{molina2019pad}, following the dense layer that processes spectral features. PAU is a trainable activation function which models nonlinearity through a rational function:
\begin{equation}
    \text{PAU}(x) = \frac{a_0 + a_1 x + a_2 x^2 + \cdots + a_J x^J}{1 + b_1 x + b_2 x^2 + \cdots + b_K x^K}
    \label{eq:pau}
\end{equation}
The coefficients $a_j$ and $b_k$ are trainable parameters. This formulation is consistent with the Padé approximant describing the weight–vibration relationship in Equation~\ref{eq:freq_solu}. In the theoretical analysis, these parameters also account for confounding structural properties, such as bed stiffness, Poisson’s ratio, thickness, size, and boundary conditions. By modeling such structural influences as trainable parameters, the physics-informed network achieves greater interpretability and improved generalizability. The polynomial orders $J$ and $K$ are determined based on the structural vibration complexity. In most scenarios, dominant vibration modes are typically fewer than five. Thus, we set $J = 5$ and $K=4$. For more complex structures with significantly higher-order modes, $J$ and $K$ can be increased accordingly as hyperparameters. 
Because the PAU structure aligns with the theoretical form of the physical relationship, it enables efficient modeling using fewer parameters, thereby improving data efficiency and generalizability by embedding domain knowledge into the learning process.

However, the PAU method is prone to instability during training because of the poles in the activation function. 
These poles can lead to erratic gradients and convergence to poor local minima. To mitigate this issue, we adopt a strategy of repeated training: we train the model independently for $N$ times and select the model with the lowest loss on the validation set for final evaluation. This approach leverages the stochastic nature of neural network training, which includes random weight initialization and the inherent randomness in batch selection. This leads different runs to converge to different local minima. By selecting the model with the lowest validation loss among the $N$ models, we reduce the risk of suboptimal solutions and improve overall stability.

In addition to the transfer function, body height is incorporated into the neural network via a physics-informed learnable sine activation function to improve weight estimation. In clinical emergency practice, length-based methods such as the Broselow tape estimate patient weight from body height measurement, demonstrating the feasibility of height acquisition in healthcare scenarios. In MelodyBedScale, height is used not only for its correlation with body weight but also for its effect on the body–bed contact area. The bed-body contact area is a confounding factor influencing the vibration transfer function spectrum through the term $C_3$, defined as the integral of the squared mode shape function over the contact area (see Section~\ref{sec:theory}). Incorporating height enables the model to disentangle the contributions of contact area from body weight, enhancing estimation accuracy. Because structural mode shapes are typically expressed in trigonometric form, we introduce a learnable sine activation of the form $\sin(\nu x)$, where $\nu$ is trainable and $x$ is the input to the activation function. This theoretically motivated activation function encodes the height information more effectively than conventional activation functions, as it aligns the network representation with the underlying physics of structural dynamics.

The outputs from the branches processing the transfer-function spectra and height are concatenated and fed into two dense layers to estimate body weight. Our physics-informed regression model embeds structural dynamics theory into the neural network via physics-informed activation functions, enabling data-efficient training, improved generalization, and more interpretable representations of the underlying physics.
\section{Real-World Human Evaluation}
\label{sec:evaluation}

To evaluate MelodyBedScale, we conducted experiments on two bed types: (1) a wooden bed with a single deck board (referred to as the wooden bed) and (2) a standard steel hospital bed with four deck boards (referred to as the steel bed). Using leave-one-person-out validation, MelodyBedScale achieved average estimation errors of 1.55 kg (2.37$\%$) on the wooden bed and 4.36 kg (6.38$\%$) on the steel hospital bed. All procedures involving human subjects were approved by the Institutional Review Board (IRB).

\subsection{Experiment Setup}
\begin{figure}[t]
    \centering
    \includegraphics[width=\linewidth]{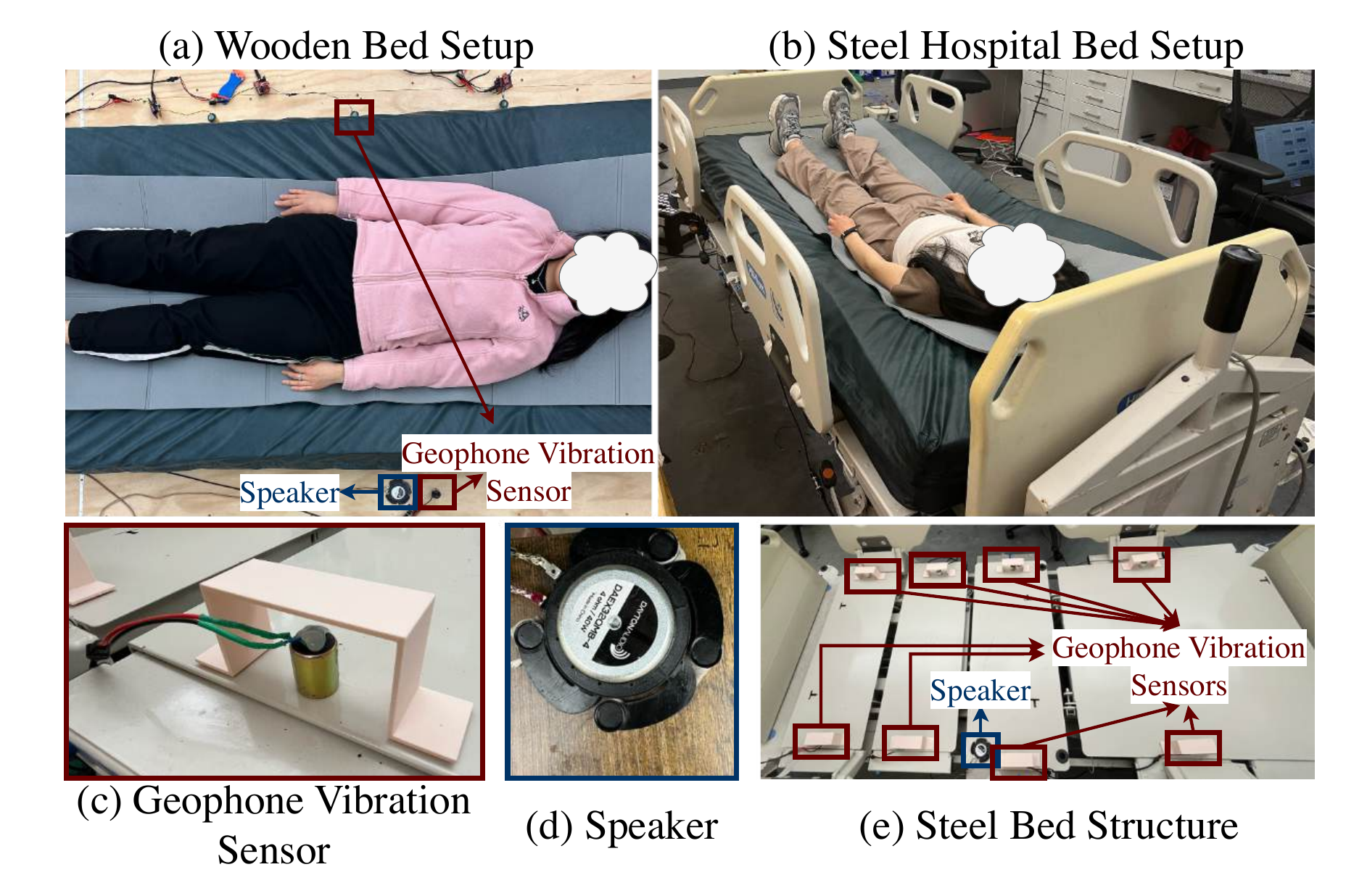}
    \caption{Experimental setups for human body weight estimation using bed structural vibrations.
(a) Wooden bed setup.
(b) Steel hospital bed setup.
(c) Geophone vibration sensor for capturing structural vibrations.
(d) Speaker used to induce bed vibrations.
(e) Steel bed structure with sensor and speaker configurations beneath the mattress. }
    \label{fig:experiment}
\end{figure}

Human experiments were conducted on two bed types with 11 participants each, collecting a total of 3,960 10-second music-induced bed vibration signal segments. The wooden bed, with dimensions of 2.44 m × 1.22 m, was supported by a single deck board beneath the mattress (see Figure~\ref{fig:experiment}a). The steel hospital bed measured 2.31 m × 0.91 m and was supported by four deck boards (see Figure~\ref{fig:experiment}b, e). In both setups, a mattress was placed on the deck boards, and participants lay on the mattress during the experiments. Bed vibrations were induced using a Dayton Audio DAEX32QMB-4 exciter speaker (see Figure~\ref{fig:experiment}d), mounted at the right-middle of the wooden bed and on the second plate of the steel bed. In both cases, the speaker was positioned near the bed’s midpoint to facilitate vibration propagation and reduce signal-to-noise ratio (SNR) decay. Music-induced bed vibrations were collected using SM-24 geophone sensors attached to the deck boards (Figure~\ref{fig:experiment}c). For the wooden bed, two sensors were employed: one near the speaker to capture the excitation signal and another on the opposite side to capture vibrations propagated through the bed-body system. For the steel bed, eight sensors were installed (two per deck board) on opposite sides of the bed to obtain vibration transfer functions for each deck board (Figure~\ref{fig:experiment}e). All vibration data were sampled at 25,600 Hz.

\begin{figure}[t]
    \centering
    \includegraphics[width=\linewidth]{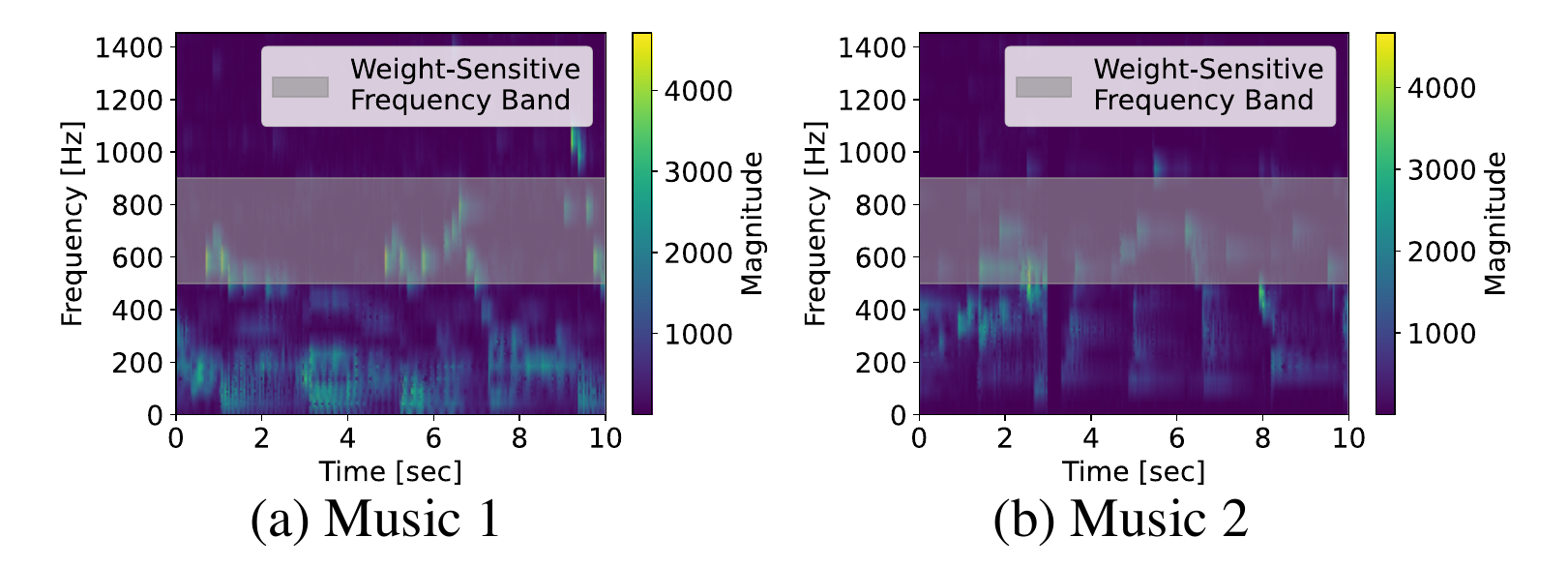} 
    \caption{Short-Time Fourier Transform (STFT) spectrograms of two musical excitations used for bed vibration induction: (a) Music 1 and (b) Music 2. }
    \label{fig:music}
\end{figure}

\begin{figure}[t]
    \centering
    \includegraphics[width=0.49\textwidth]{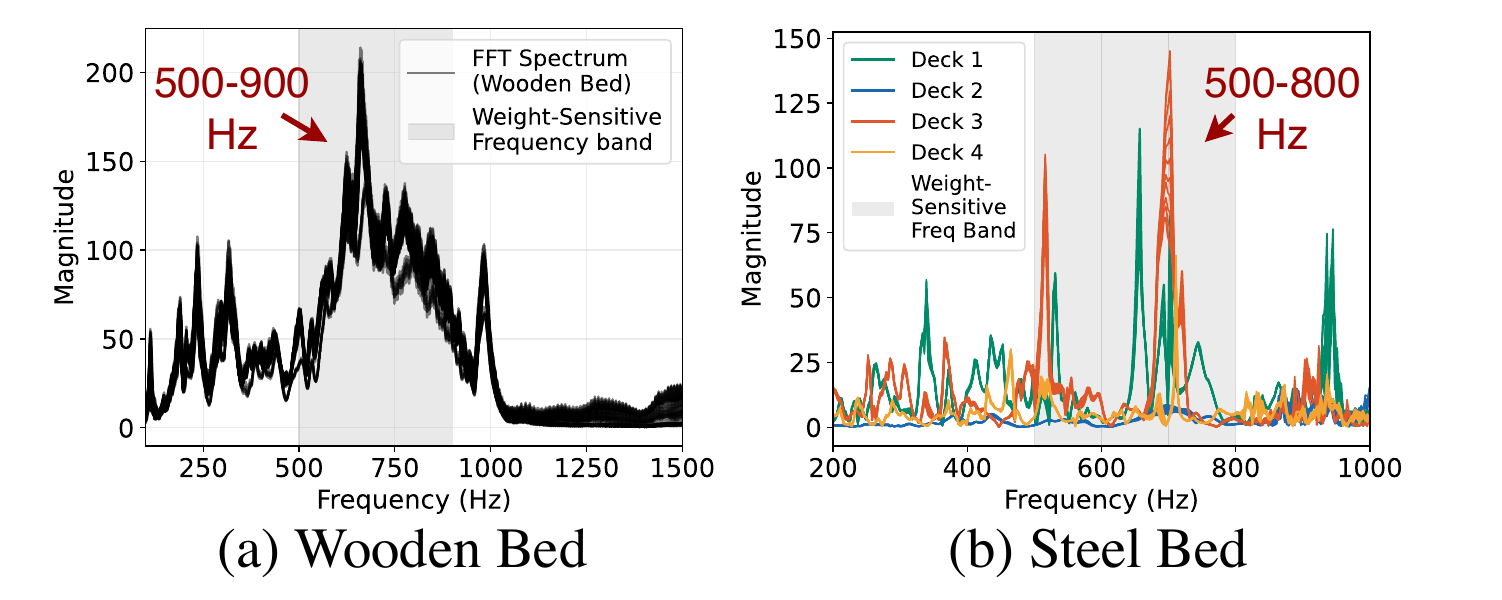} 
    \caption{Vibration spectra induced by chirp excitation for (a) the wooden bed and (b) the steel bed. }
    \label{fig:weight_sensitive}
\end{figure}

The wooden bed experiments employed two music tracks (short-time Fourier transform spectra in Figure~\ref{fig:music}), and the steel bed tests included only Music 1. Both tracks were selected for their high energy within the weight-sensitive frequency band of 500–900 Hz. The weight-sensitive frequency band was identified using a chirp signal sweeping from 10–1000 Hz with uniform energy distribution. For the wooden and steel beds, the identified weight-sensitive bands were 500–900 Hz and 500–800 Hz, respectively (see Figure~\ref{fig:weight_sensitive}).

Participants with diverse body weights were recruited for both bed types. For the wooden bed, participant characteristics were: height 150–180 cm, weight 48.45–84.15 kg, BMI 20.04–25.97; and for the steel bed: height 160–187 cm, weight 50.45–90.40 kg, BMI 19.71–29.66. Before each trial, participants’ body weight was measured using a standard scale. Participants then lay supine on a bed, and music was played through a speaker to induce bed vibrations. Each music sample lasted 10 s, during which synchronized vibration sensors were measuring the music-induced bed vibration signals. Each music type was repeated 20 times. During music playing time, participants were allowed to make minor movements (e.g., lifting an arm) but instructed to avoid large movements such as sitting up or turning the torso. Then, to simulate varying body weights, participants held water bottles on their abdomen, each containing 1 L of water (1 kg). Starting with a 1 kg water bottle, the procedure was repeated in 1 kg increments up to 5 kg. For each added weight, the same music was replayed, and vibration data were collected. 
In total, 1,320 vibration segments were collected per bed type for each music: for each participant, 20 ten-second segments were recorded at six weight levels (self-weight plus 0–5 kg). Across both bed types, this yielded 3,960 10-second vibration signal groups, with each group corresponding to simultaneous signals from all sensors.

\subsection{Model Hyperparameter Configuration}
The implementation details and hyperparameters of the body weight regression neural network are summarized as follows. The transfer function spectra for model input were selected within the 500–900 Hz range for the wooden bed and 500-800 Hz for the steel bed, identified as the weight-sensitive frequency band (see Figure~\ref{fig:weight_sensitive}). The transfer function encoder included two hidden layers (128 and 64 units), while the height encoder comprised two hidden layers of 64 units each. Their outputs were concatenated into a 128-dimensional vector and passed through a dense network with a 128→64 linear layer (ReLU activation) followed by a 64→1 output layer. In the PAU activation unit, the numbers of $a_j$ and $b_k$ were set to 5 and 4, respectively, to capture the dominant vibration modes within the first few modes. The network contained approximately 208k trainable parameters. Training employed an L1 loss function, a learnable sine activation function initialized at frequency 1.0, and the Adam optimizer (learning rate = 0.001, weight decay = 0.0005) for 2000 epochs. StandardScaler normalization was applied independently to spectra and height inputs, and target weights were standardized to zero mean and unit standard deviation, then rescaled for evaluation. Each configuration was trained five times, and the model with the lowest validation L1 loss was selected, as detailed in Section~\ref {sec:phy_method}.

\subsection{Body Weight Estimation Results}
We evaluated MelodyBedScale using both leave-one-person-out (LOPO) and leave-one-weight-out (LOWO) cross- validation on the collected dataset. For the wooden bed, the mean absolute error (MAE) in body weight estimation was 1.55 kg (2.37 $\%$) under leave-one-person-out validation and 0.67 kg (0.98 $\%$) under leave-one-weight-out validation. For the steel hospital bed, the MAE was 4.36 kg (6.38 $\%$) for leave-one-person-out validation and 1.06 kg (1.56 $\%$) for leave-one-weight-out validation. In all cases, MelodyBedScale exceeded the clinical requirement that $>70\%$ of estimates are within 10$\%$ error and $>95\%$ within 20$\%$ error~\cite{wells2017accuracy}. Moreover, under leave-one-person-out validation, MelodyBedScale reduced estimation error by 78.26 $\%$ and 68.14 $\%$ compared with height-based estimation alone for the wooden bed and steel bed, respectively.

\begin{figure}[t]
    \centering
    \includegraphics[width=\linewidth]{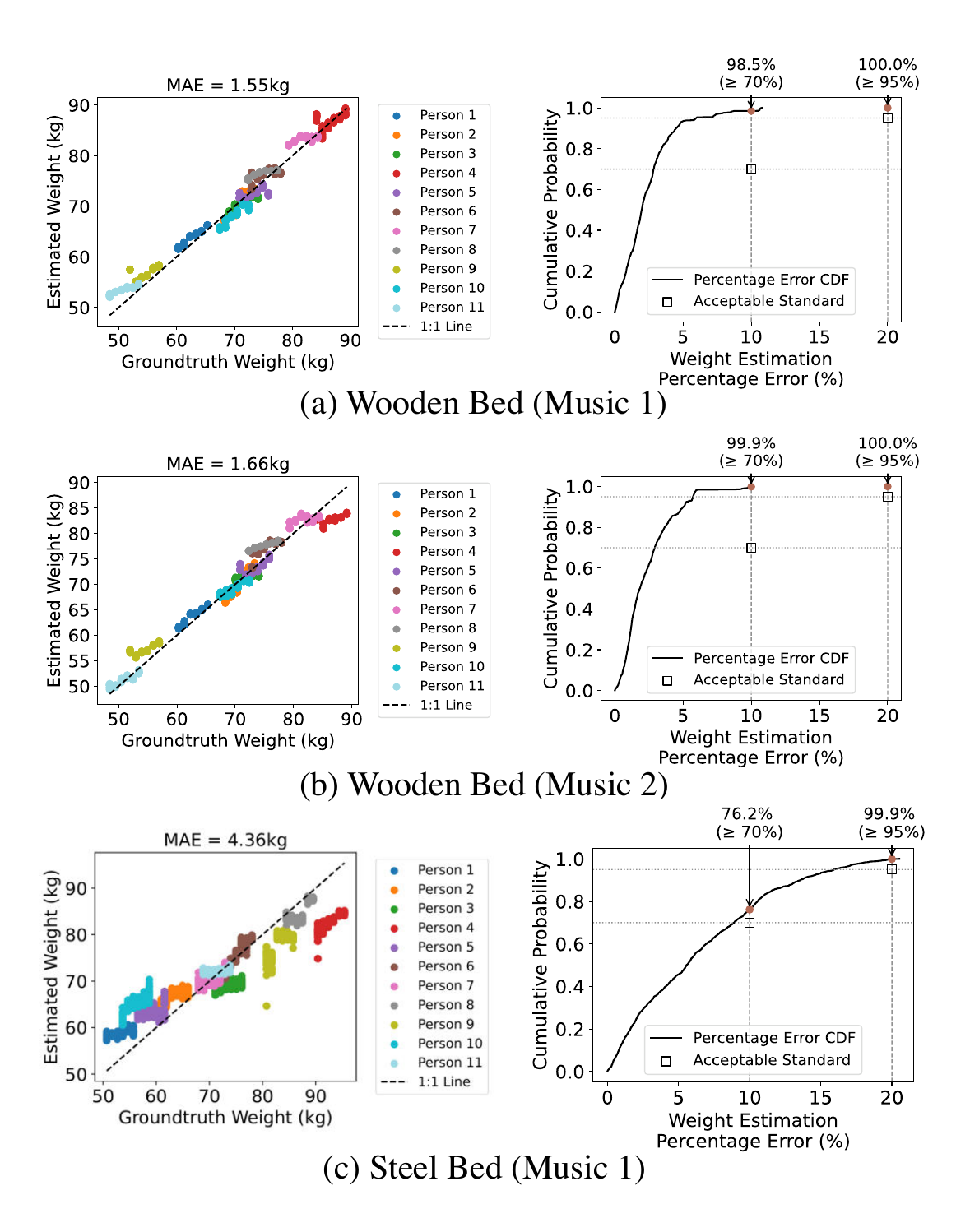} 
    \caption{Body weight estimation on (a) wooden bed with Music 1, (b) wooden bed with Music 2, and (c) steel bed with Music 1. Left: scatterplots of estimated vs. ground-truth weight (MAE = 1.55, 1.66, and 4.36 kg). Right: cumulative distribution functions (CDFs) of percentage estimation errors. }
    \label{fig:results_combined}
\end{figure}

\subsubsection{Leave-One-Person-Out Cross-Validation}

\begin{figure*}[t]
    \centering
    \includegraphics[width=\textwidth]{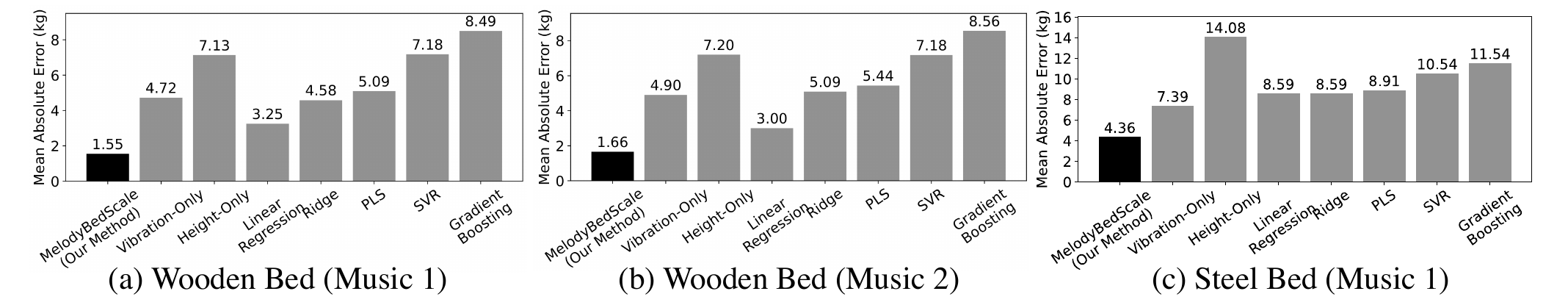} 
    \caption{Mean absolute errors of weight estimation across different methods for (a) wooden bed under Music 1, (b) wooden bed under Music 2, and (c) steel bed under Music 1.}
    \label{fig:baseline_compare}
\end{figure*}

Our system achieved mean absolute errors (MAE) of 1.55 kg (2.37$\%$) and 1.66 kg (2.36$\%$) for music 1 and music 2 on the wooden bed, and 4.36 kg (6.38$\%$) on the steel bed (see Figure~\ref{fig:results_combined}). In the leave-one-person-out (LOPO) validation, data from one participant was reserved for testing, while data from the remaining 10 participants was split into 80$\%$ training and 20$\%$ validation. This procedure was repeated for each participant, yielding 11 models in total, and the reported MAE reflects the average across these models. The scatterplots (Figure~\ref{fig:results_combined}, left) illustrate strong alignment between estimated and ground-truth weights, closely following the 1:1 reference line. Pearson correlation coefficients between estimated and ground-truth weights were 0.986, 0.981, and 0.950 for the wooden bed with music 1, the wooden bed with music 2, and the steel bed, respectively. The cumulative distribution functions (CDFs) (Figure~\ref{fig:results_combined}, right) present the distribution of percentage errors, calculated as absolute error divided by ground truth body weight. For the wooden bed, 98.5$\%$ (music 1) and 99.9$\%$ (music 2) of samples fell within 10$\%$ estimation error, substantially outperforming the clinically acceptable standard (square markers shown in Figure~\ref{fig:results_combined}, right). On the steel bed, MelodyBedScale achieved an MAE of 4.36 kg (6.38$\%$), with 76.2$\%$ of estimates within 10$\%$ error and 99.9$\%$ within 20$\%$ error, satisfying clinical standards. 
The wooden bed generally outperformed the steel bed with lower MAE values. This performance difference is likely due to the structural complexity: the wooden bed had a simple single-deck design, whereas the steel hospital bed incorporates multiple mechanical components underneath for height adjustment and mobility, including four deck boards and complex wheel connections, which increase the structural transfer function’s complexity and make it more difficult to model accurately.

We compared our method against baseline models using either vibration data or height information only, as well as several standard regressors, including linear regression, ridge regression, partial least squares (PLS), support vector regression (SVR), and gradient boosting. In all three scenarios, our method consistently achieved the lowest MAE, outperforming all baselines (see Figure~\ref{fig:baseline_compare}).

\begin{figure}[t]
    \centering
    \includegraphics[width=0.9\linewidth]{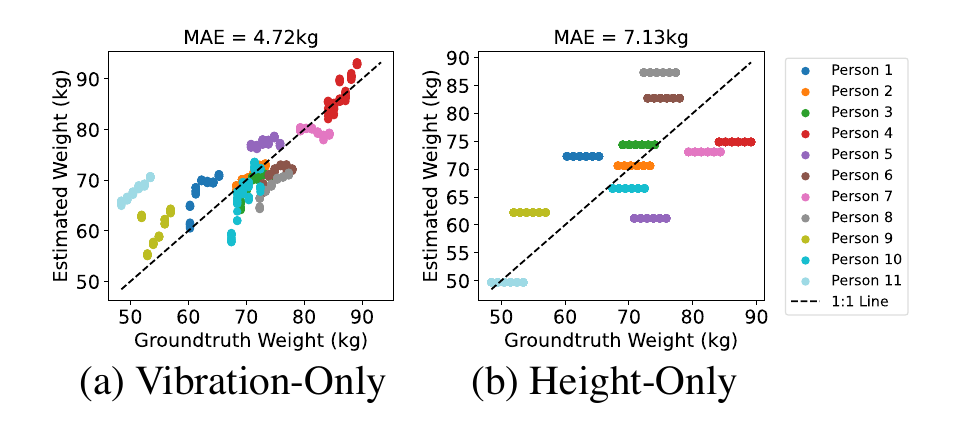} 
    \caption{Ablation study with (a) vibration-only and (b) height-only information (wooden bed, Music 1).}
    \label{fig:scatter_compare}
\end{figure}
We evaluated the individual contributions of vibration and height features through ablation studies, using only vibration data or only height data for weight estimation on the wooden bed music 1 dataset. Figure~\ref{fig:scatter_compare} presents scatterplots comparing estimated weights against ground truth for both scenarios. The vibration-only model achieved a mean absolute error (MAE) of 4.72 kg, representing a 3.05x increase over our full model. This approach exhibited poor performance for lighter individuals, likely due to the dataset's limited representation of lower-weight subjects (45-65 kg) and the model's reduced accuracy out of training weight distribution. The height-only model produced an MAE of 7.13 kg (4.6x higher than ours). This demonstrates its inability to account for body-shape variations when estimating weight from height alone. As illustrated in Figure~\ref{fig:scatter_compare}b, individuals with similar weights ($\approx$70 kg) displayed substantial height differences, causing the significant estimation errors of the height-only model.

\begin{figure}[t]
    \centering
    \includegraphics[width=\linewidth]{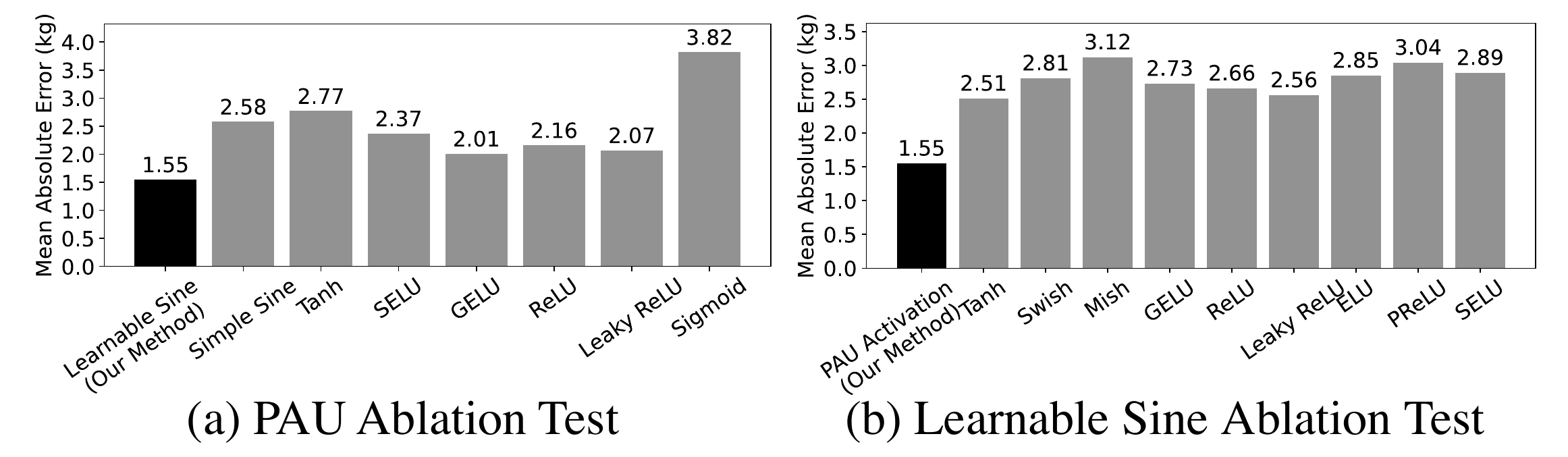} 
    \caption{Ablation study results for (a) PAU and (b) learnable sine (wooden bed, Music 1). }
    \label{fig:act_ablation}
\end{figure}

To evaluate the effectiveness of the physics-informed activation functions (PAU and learnable sine), we conducted ablation experiments on the wooden bed dataset with music 1. Each activation was replaced with commonly used alternatives while keeping all other network components unchanged. For the PAU ablation, we substituted PAU with \texttt{tanh}, \texttt{swish}, \texttt{mish}, \texttt{gelu}, \texttt{relu}, \texttt{leakyrelu}, \texttt{elu}, \texttt{PReLU}, and \texttt{SeLU}. Our method with PAU achieved the lowest MAE, indicating the effectiveness of PAU (see Figure~\ref{fig:act_ablation}a). For the learnable sine ablation, we replaced it with simple sine, \texttt{tanh}, \texttt{selu}, \texttt{gelu}, \texttt{relu}, \texttt{leakyrelu}, and \texttt{sigmoid}. Compared to other activation functions, the learnable sine yielded the lowest MAE, demonstrating its effectiveness (see Figure~\ref{fig:act_ablation}b).

\subsubsection{Leave-One-Weight-Out Cross-Validation}
\begin{figure}[htbp]
    \centering
    \includegraphics[width=\linewidth]{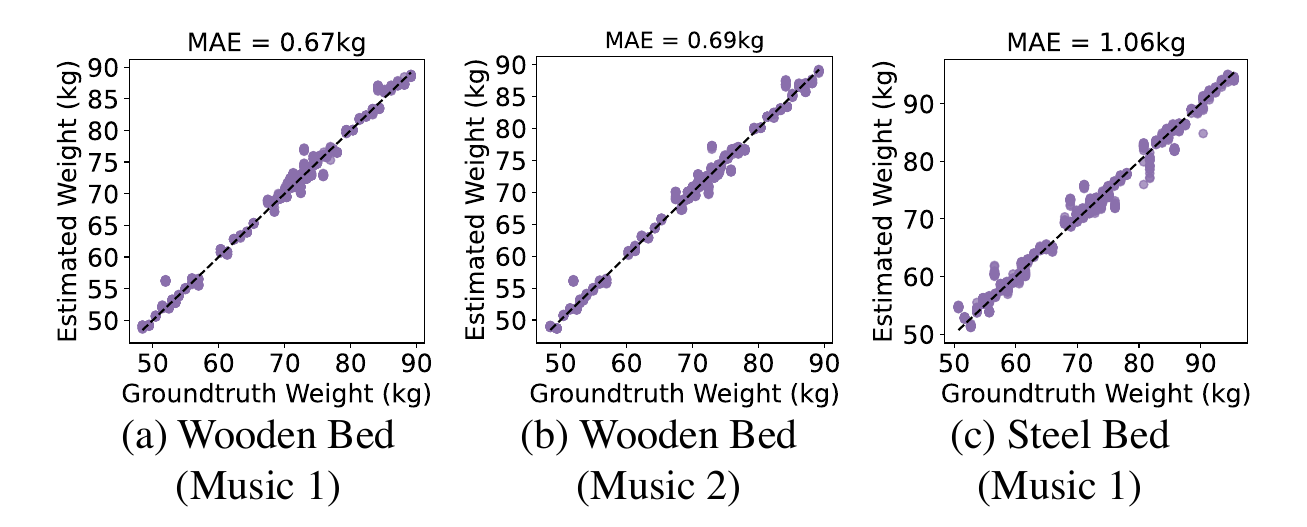} 
    \caption{Leave-one-weight-out cross-validation results for (a) a wooden bed with Music 1, (b) a wooden bed with Music 2, and (c) a steel bed with Music 1.}
    \label{fig:lowo}
\end{figure}
A leave-one-weight-out (LOWO) cross-validation was implemented for the body weight estimation task. 
In each LOWO iteration, data corresponding to one specific weight for a given participant were used as the test set, while data from all other weights (including other participants’ data and the remaining weights for the same participant) were used for training and validation. The results for the wooden bed and steel hospital bed are shown in Figures~\ref{fig:lowo}. On the wooden bed, the mean absolute error (MAE) was 0.67 kg (0.98 $\%$) for Music 1 and 0.69 kg (1.01 $\%$) for Music 2. On the steel bed, the MAE was 1.06 kg (1.56 $\%$). These results outperformed the leave-one-person-out validation because the training set in LOWO includes data from the same participant, preserving the same contact area and providing more weight distributions similar to the test set, thereby improving estimation accuracy.


\subsection{Model Robustness Test}
In this subsection, we evaluated the robustness of MelodyBedScale under varying conditions, including different individuals, sound volume changes, sensor noise, the material of added weight, and lying postures. The results indicated that the system exhibited robustness against all these factors.

\subsubsection{Robustness Across Different Individuals}
\begin{figure}[t]
    \centering
    \includegraphics[width=\linewidth]{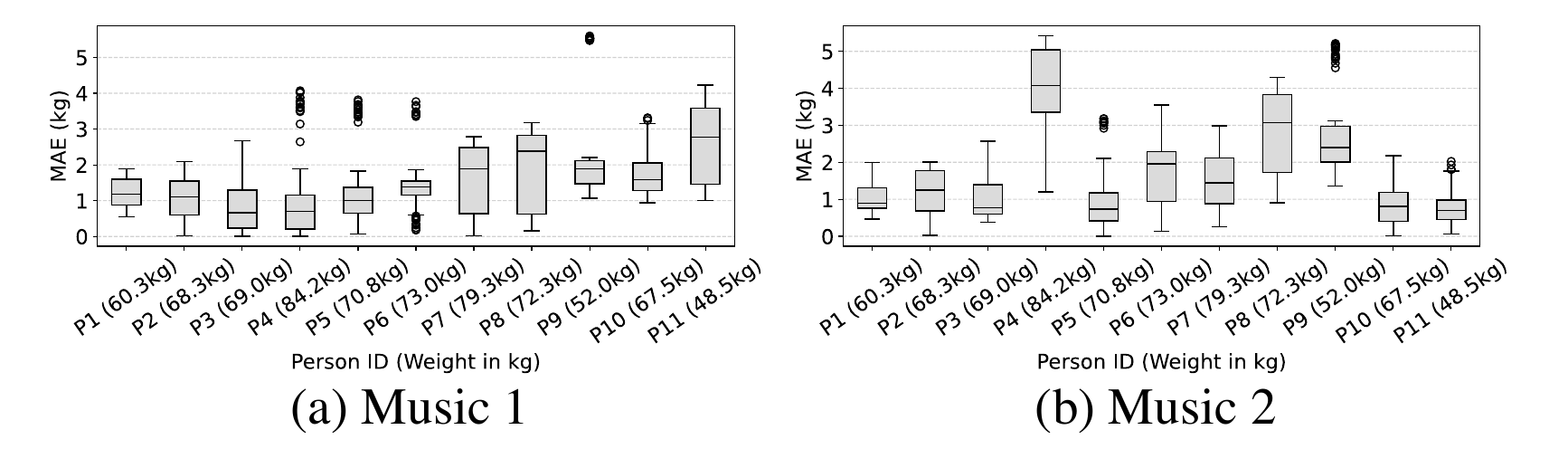} 
    \caption{Distribution of mean absolute error (MAE) in weight estimation across participants on a wooden bed for (a) Music 1 and (b) Music 2.}
    \label{fig:person_diff}
\end{figure}
We evaluated the robustness of the MelodyBedScale system across different individuals by analyzing the weight estimation error distributions using the wooden bed dataset. Figure~\ref{fig:person_diff} shows boxplots of the MAEs for each individual. Despite the diverse heights and BMI profiles among these participants, the system performance was generally consistent, with mean absolute errors below 4.5 kg for all individuals. 
For Music 2, the error was largest for Participant 4 (MAE = 3.82 kg), the heaviest individual in the dataset (84.2 kg).
A possible explanation is that excluding the heaviest individual during training in LOPO reduces the coverage of the weight range in the dataset, leading to degraded accuracy for out-of-distribution weight values. To this end, for real‑world deployments, it is necessary to cover the lower and upper bounds of the expected weight range in the training set.

\subsubsection{Robustness to Sound Volume of Music Excitation}

\begin{figure}[t]
    \centering
    \includegraphics[width=\linewidth]{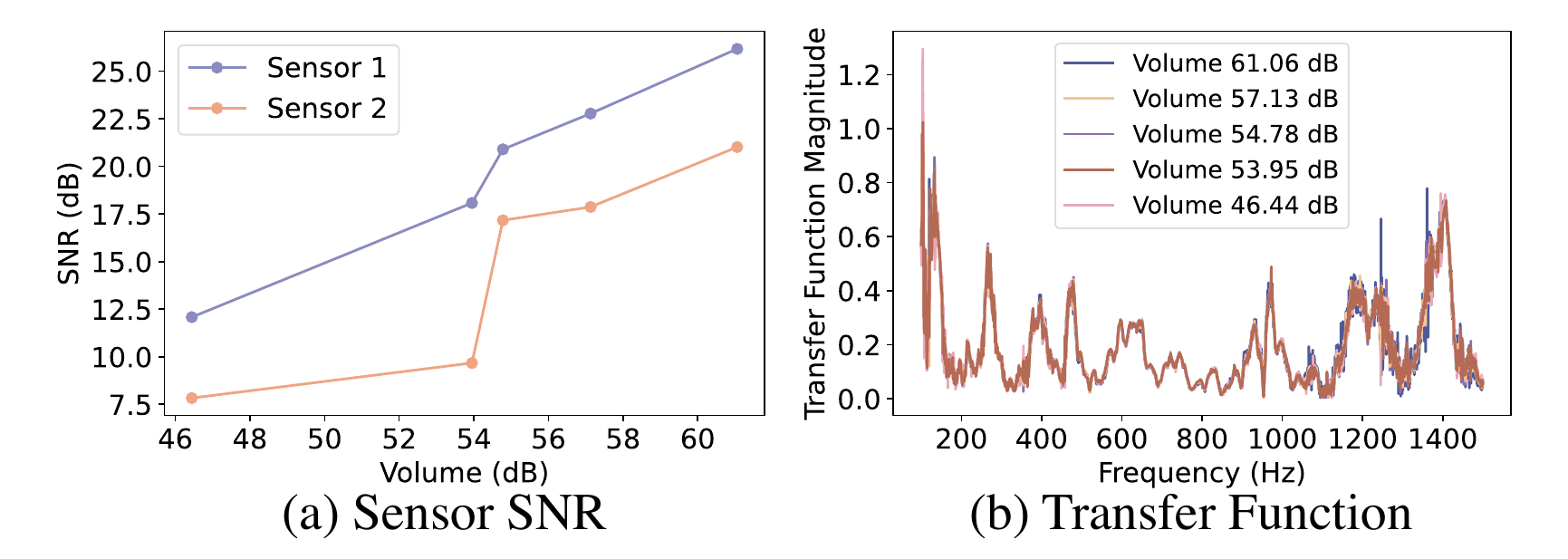} 
    \caption{(a) Signal-to-noise ratio (SNR) of two sensors under varying music excitation volume levels. Acceptable SNR (>10 dB) at volumes of 54.78 dB. (b) Transfer function spectra across excitation volumes (46–61 dB). }
    \label{fig:sound_volume}
\end{figure}

We evaluated system robustness to varying music volumes by gradually reducing sound levels and analyzing the corresponding changes in the vibration transfer function, using a wooden bed as an example.
During our experiments, music at approximately 60 dB was used to induce structural bed vibrations, a level acceptable in clinical settings. Music sound levels were measured with the Phyphox app on an iPhone, and sensor signal-to-noise ratio (SNR) was evaluated across varying sound volumes. As shown in Figure~\ref{fig:sound_volume}a, sensor SNR generally increased with volume. An SNR of 10 dB, which is typically considered acceptable for sensing applications, corresponds to a music volume of approximately 54 dB, which is comparable to normal conversation levels. SNR was calculated as $10 \log_{10}(signal\ energy /noise\ energy)$, where noise energy corresponds to environmental noise without music playing. We analyzed transfer function spectra between two vibration sensors under different sound volumes (Figure~\ref{fig:sound_volume}b) with the wooden bed. The transfer function spectra showed only minor variations, indicating the robustness of the system to reduced music excitation volumes.

\subsubsection{Robustness to Sensor Noise}
\begin{figure}[t]
    \centering
    \includegraphics[width=\linewidth]{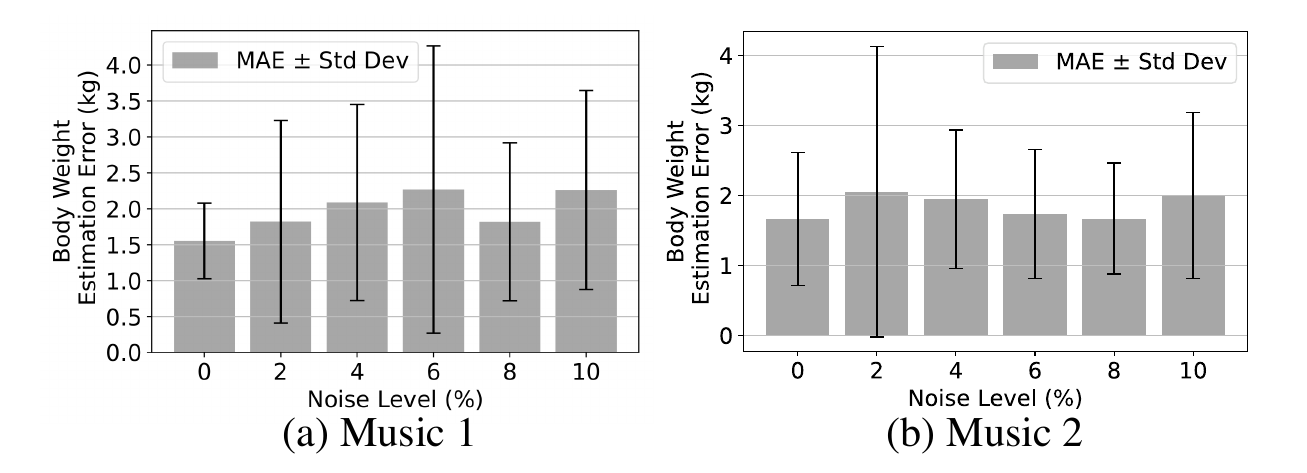} 
    \caption{Mean absolute error (MAE) and standard deviation of body weight estimation across varying levels of Gaussian noise added to sensor signals for (a) Music 1 and (b) Music 2 (wooden bed). }
    \label{fig:noise}
\end{figure}

To evaluate the robustness of MelodyBedScale to sensor noise, Gaussian noise was added to the vibration signals before computing the vibration transfer function. Noise levels were set at 2$\%$, 4$\%$, 6$\%$, 8$\%$, and 10$\%$ of the vibration signal amplitude. Evaluation was performed on the wooden bed dataset with both music 1 and 2. Figure~\ref{fig:noise} shows the mean absolute error (MAE) for weight estimation, with standard deviation bars indicating the error distribution under different noise levels. As shown, increasing noise resulted in only minor increases in MAE, indicating that MelodyBedScale maintains robust performance in the presence of additive Gaussian noise.

\subsubsection{Robustness to Added Weight Material}
The material of the added weight does not influence the performance of the MelodyBedScale system. In our experiments, water bottles were used as added weight to simulate increases in human body weight. We assumed that differences in vibration signals are primarily attributable to the magnitude of the added weight and are invariant to the material of the added weight.
To validate this assumption, we collected data from the same subject (body weight: 52.80 kg) while holding water bottles and while holding steel dumbbells of varying masses. With dumbbells, the subject’s total weight was 54.15, 55.50, 56.65, 57.70, and 58.80 kg. A model was trained on data collected with water bottles (total weight range: 52.80–63.45 kg) and tested on data with steel dumbbells. The model achieved a mean absolute error of 0.85 kg. The low estimation error confirms that the performance of MelodyBedScale is robust to variations in the material of the added weight.
%

\begin{figure}[t]
    \centering
    \includegraphics[width=\linewidth]{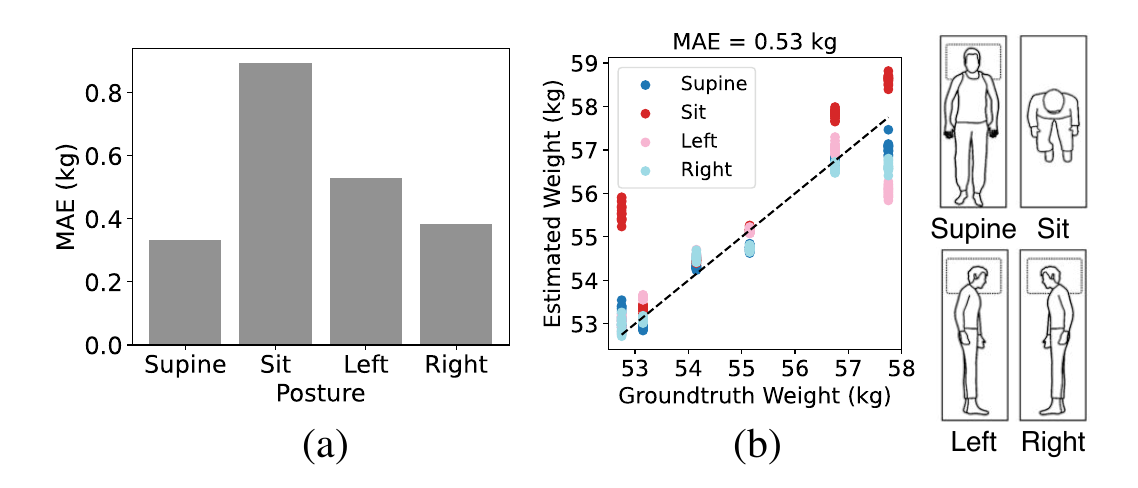} 
    \caption{Leave-one-posture-out cross-validation results. (a) MAE by posture. (b) Scatterplot of estimated versus ground-truth weight across postures.}
    \label{fig:posture}
\end{figure}
\subsubsection{Robustness to Lying Postures}
We evaluated system performance across multiple postures for the robustness test. Experiments were conducted with one participant across four postures: supine, sitting, left lateral (referred to as left), and right lateral (referred to as right). The leave-one-posture-out cross-validation was conducted, where each posture was excluded in turn as the test set while the remaining three were used for training. As shown in Figure~\ref{fig:posture}, the system achieved a mean absolute error (MAE) of 0.53 kg overall, with posture-specific errors of 0.33 kg (supine), 0.89 kg (sitting), 0.53 kg (left), and 0.38 kg (right). The sitting posture yielded the largest error, likely due to its distinct bed–body contact area compared to the other three postures, which share similar contact characteristics. Nevertheless, the system demonstrated robust performance across all postures.

\subsection{Validation of Weight-Sensitive Frequency Bands}
In this subsection, we validate that the identified weight-sensitive bands enable more effective music excitation and lead to improved model performance through more effective feature extraction.

\subsubsection{Validation of Music Excitation in Weight-Sensitive Frequency Bands}
To validate that higher energy in weight-sensitive frequency bands enhances model performance, we analyzed the relationship between band-specific energy distributions of music-induced bed vibrations and mean absolute error (MAE) in body weight estimation. The analysis was conducted using data collected from the wooden bed as a representative example. In this analysis, music 1 and 2 were segmented into two-second windows with one-second overlap (38 music windows in total). For each window, the transfer function was computed and used as input to the physics-informed model for body weight estimation. We then analyzed the relationship between frequency band energy distributions of the music excitations and the corresponding body weight estimation errors to identify excitation patterns that improve estimation accuracy.
For each music excitation window, spectral energy was calculated within consecutive 200 Hz frequency bands with 100 Hz overlap (0–200 Hz, 100–300 Hz, 200–400 Hz, etc.) up to 1500 Hz. A random forest model was trained using logarithmic band energies as input to predict the MelodyBedScale MAE for body weight estimation. This model achieved an average error of 0.7 kg under 10-fold cross-validation. This model predicted the MAE of MelodyBedScale weight estimation given the music excitation used to induce bed vibrations.

\begin{figure}[t]
    \centering
    \includegraphics[width=0.7\linewidth]{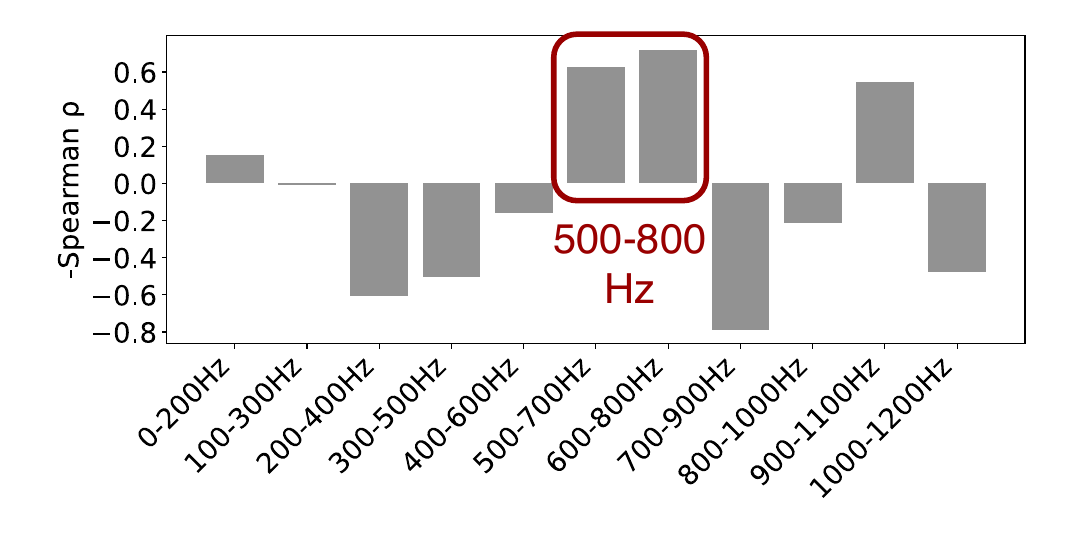} 
    \caption{Negative Spearman correlation ($\rho$) between the log energy of each frequency band and SHAP feature importance values of weight estimation MAE.}
    \label{fig:correlation}
\end{figure}

The SHAP (SHapley Additive exPlanations) method~\cite{sundararajan2020many} was applied to quantify the contribution of each frequency-band energy to the random forest model’s mean absolute error (MAE) in body weight estimation. 
In our analysis, the SHAP value represented how a given frequency band's log-energy affects the predicted MAE for body weight estimation. Positive correlations between SHAP values and log band energy indicate that higher energy values in this frequency band increase estimation error, whereas negative correlations imply error reduction. Consequently, more effective frequency bands should exhibit stronger negative correlations between SHAP values and log-energy values.
Figure~\ref{fig:correlation} presents the negative Spearman correlation coefficients between SHAP values and log-energy for each frequency band. The 500–700 Hz and 600–800 Hz bands demonstrated the strongest negative correlations with coefficients of –0.628 and –0.719, respectively. These findings align with the weight-sensitive modal natural frequencies (500–900 Hz). 

\subsubsection{Validation of Vibration Transfer Function Feature Selection in Weight-Sensitive Frequency Bands}
\begin{figure}[t]
    \centering
    \includegraphics[width=\linewidth]{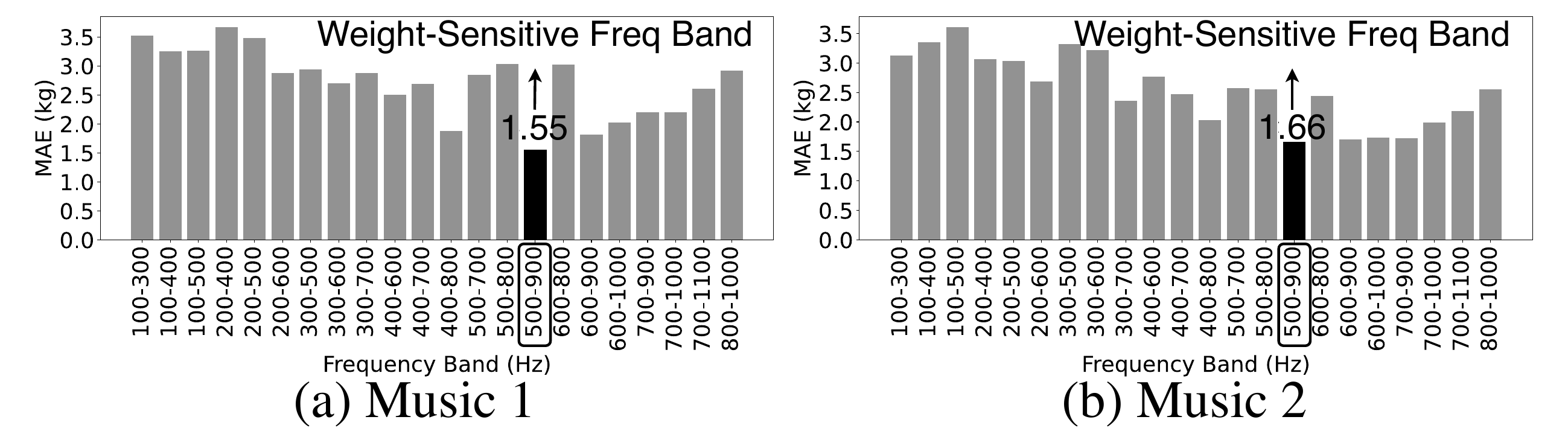} 
    \caption{Model performance comparison using transfer function spectra from different frequency bands: (a) Music 1 and (b) Music 2 (wooden bed). }
    \label{fig:feature_selec}
\end{figure}

We validate the effectiveness of selecting the weight-sensitive frequency band of the transfer function as input features for the regression model by comparing performance against alternative band selections.
In our experiments, the vibration transfer function within 500–900 Hz for the wooden bed and 500–800 Hz for the steel bed was selected as the input feature for the physics-informed body weight regression model, as this band was identified as weight-sensitive in previous analysis (Section~\ref{sec:freq_band_sele}). To validate its effectiveness, we compared model performance using features from alternative frequency bands on the wooden bed, with results summarized in Figure~\ref{fig:feature_selec}. As shown, the 500–900 Hz band yielded the lowest MAE, thereby confirming the validity of selecting the weight-sensitive frequency band from the transfer function spectra as input to the body weight regression model.


\section{Related Works}
\label{sec:related_works}

\textbf{Patient Weight Measurement Methods:}
Current methods for measuring body weight in immobilized patients include visual assessment, length-based tapes~\cite{lubitz1988rapid,wells2013pawper,abdel2012improved}, transfer-based weighing scales~\cite{healthometer2019accuracy,charder2024ms6001}, bed-integrated load cells~\cite{vpgforcesensors2024hospital,ishikawa2022long,harrington2021passive}, and pressure mats~\cite{wu2023massnet,kim2020prediction,davoodnia2023deep}. These approaches are often limited by inaccuracy~\cite{anglemyer2004accuracy,wells2017accuracy}, the need for patient or bed movement, which is labor-intensive and time-consuming, or by requiring patient removal for calibration or installation. In contrast, MelodyBedScale provides a rapid, non-invasive, and accurate alternative for body weight estimation.

\textbf{Vibration-Based Weight Estimation Methods:} 
Prior studies have leveraged structural vibrations to estimate the weight of vehicles, livestock, and other objects.
Vehicle weigh-in-motion systems analyze vehicle-induced road vibrations for weight estimation~\cite{bajwa2017development}. However, this passive vibration sensing system does not work for static objects. Active vibration sensing methods for weight estimation of livestock and objects~\cite{codling2021masshog, zhang2025wevibe, zhang2020vibroscale} either fail to generalize to unseen weights, rely on sweep-frequency excitation, or require zero-load reference signals. Thus, they are unsuitable for clinical patient weight measurement on beds. Compared to the existing methods, MelodyBedScale addresses unique clinical challenges, including patient immobility constraints and acoustic restrictions in hospital environments.

\textbf{Trainable Activation Functions:}
Incorporating physics-inspired learnable activation functions into neural networks has proven effective for improving efficiency and generalizability~\cite{abbasi2024physical}. Trainable activation functions allow models to capture complex patterns more effectively~\cite{apicella2021survey}. Examples include parameterized variants of standard functions~\cite{he2015delving,ramachandran2017swish,trottier2017parametric} and methods that learn combinations of different activations~\cite{sutfeld2020adaptive,ma2021activate,chelly2024trainable}, which offer greater flexibility but often require larger datasets. Our trainable activation function selection, PAU~\cite{molina2019pad}, is guided by structural dynamics theory to achieve data-efficient training with stronger generalization on limited data.

\section{Conclusions and Future Works}
\label{sec:conclusion}
In this paper, we present MelodyBedScale, a novel system for estimating human body weight through music-induced bed vibrations. The system first identifies weight-sensitive frequency bands and then composes clinically acceptable music with high energy in these bands to induce bed vibrations. We formulate the complex and indirect weight-vibration relationship through theoretical structural dynamics analysis and incorporate the results into physics-informed regression models by shaping activation functions accordingly. This enables a data-efficient, interpretable, and generalizable weight estimation model. 
We evaluated MelodyBedScale on two bed types with 11 participants each, achieving mean absolute errors of 1.55 kg (2.37 $\%$) and 4.36 kg (6.38 $\%$) under leave-one-person-out cross-validation. In both cases, our results satisfied clinically acceptable standards.

In future work, we aim to address the confounding effects of bed–body contact area on vibration responses by incorporating thermal imaging to infer the lying postures and positions. In addition, we plan to extend our experiments to pediatric patients in emergency care settings.
\begin{acks}

This project is funded by the Stanford Blume Fellowship and the Stanford CEE-PhD Fellowship.

\end{acks}

\bibliographystyle{ACM-Reference-Format}
\bibliography{01_main}


\begin{thebibliography}{41}


\ifx \showCODEN    \undefined \def \showCODEN     #1{\unskip}     \fi
\ifx \showISBNx    \undefined \def \showISBNx     #1{\unskip}     \fi
\ifx \showISBNxiii \undefined \def \showISBNxiii  #1{\unskip}     \fi
\ifx \showISSN     \undefined \def \showISSN      #1{\unskip}     \fi
\ifx \showLCCN     \undefined \def \showLCCN      #1{\unskip}     \fi
\ifx \shownote     \undefined \def \shownote      #1{#1}          \fi
\ifx \showarticletitle \undefined \def \showarticletitle #1{#1}   \fi
\ifx \showURL      \undefined \def \showURL       {\relax}        \fi
\providecommand\bibfield[2]{#2}
\providecommand\bibinfo[2]{#2}
\providecommand\natexlab[1]{#1}
\providecommand\showeprint[2][]{arXiv:#2}

\bibitem[Abbasi and Andersen(2024)]%
        {abbasi2024physical}
\bibfield{author}{\bibinfo{person}{Jassem Abbasi} {and} \bibinfo{person}{P{\aa}l~{\O}steb{\o} Andersen}.} \bibinfo{year}{2024}\natexlab{}.
\newblock \showarticletitle{Physical activation functions (PAFs): An approach for more efficient induction of physics into physics-informed neural networks (PINNs)}.
\newblock \bibinfo{journal}{\emph{Neurocomputing}}  \bibinfo{volume}{608} (\bibinfo{year}{2024}), \bibinfo{pages}{128352}.
\newblock


\bibitem[Abdel-Rahman and Ridge(2012)]%
        {abdel2012improved}
\bibfield{author}{\bibinfo{person}{Susan~M Abdel-Rahman} {and} \bibinfo{person}{Anna~L Ridge}.} \bibinfo{year}{2012}\natexlab{}.
\newblock \showarticletitle{An improved pediatric weight estimation strategy}.
\newblock \bibinfo{journal}{\emph{Open Med Dev J}} \bibinfo{volume}{4}, \bibinfo{number}{4} (\bibinfo{year}{2012}), \bibinfo{pages}{87--97}.
\newblock


\bibitem[Anglemyer et~al\mbox{.}(2004)]%
        {anglemyer2004accuracy}
\bibfield{author}{\bibinfo{person}{Bradley~L Anglemyer}, \bibinfo{person}{Chris Hernandez}, \bibinfo{person}{Jane~H Brice}, {and} \bibinfo{person}{Bin Zou}.} \bibinfo{year}{2004}\natexlab{}.
\newblock \showarticletitle{The accuracy of visual estimation of body weight in the ED}.
\newblock \bibinfo{journal}{\emph{The American journal of emergency medicine}} \bibinfo{volume}{22}, \bibinfo{number}{7} (\bibinfo{year}{2004}), \bibinfo{pages}{526--529}.
\newblock


\bibitem[Apicella et~al\mbox{.}(2021)]%
        {apicella2021survey}
\bibfield{author}{\bibinfo{person}{Andrea Apicella}, \bibinfo{person}{Francesco Donnarumma}, \bibinfo{person}{Francesco Isgr{\`o}}, {and} \bibinfo{person}{Roberto Prevete}.} \bibinfo{year}{2021}\natexlab{}.
\newblock \showarticletitle{A survey on modern trainable activation functions}.
\newblock \bibinfo{journal}{\emph{Neural Networks}}  \bibinfo{volume}{138} (\bibinfo{year}{2021}), \bibinfo{pages}{14--32}.
\newblock


\bibitem[Authority(2009)]%
        {authority2009medication}
\bibfield{author}{\bibinfo{person}{Pennsylvania Patient~Safety Authority}.} \bibinfo{year}{2009}\natexlab{}.
\newblock \showarticletitle{Medication errors: significance of accurate patient weights}.
\newblock \bibinfo{journal}{\emph{Pa Patient Saf Advis}} \bibinfo{volume}{6}, \bibinfo{number}{1} (\bibinfo{year}{2009}), \bibinfo{pages}{10--15}.
\newblock


\bibitem[Bajwa et~al\mbox{.}(2017)]%
        {bajwa2017development}
\bibfield{author}{\bibinfo{person}{Ravneet Bajwa}, \bibinfo{person}{Erdem Coleri}, \bibinfo{person}{Ram Rajagopal}, \bibinfo{person}{Pravin Varaiya}, {and} \bibinfo{person}{Christopher Flores}.} \bibinfo{year}{2017}\natexlab{}.
\newblock \showarticletitle{Development of a cost-effective wireless vibration weigh-in-motion system to estimate axle weights of trucks}.
\newblock \bibinfo{journal}{\emph{Computer-Aided Civil and Infrastructure Engineering}} \bibinfo{volume}{32}, \bibinfo{number}{6} (\bibinfo{year}{2017}), \bibinfo{pages}{443--457}.
\newblock


\bibitem[Bendat and Piersol(1986)]%
        {bendat1986random}
\bibfield{author}{\bibinfo{person}{JS Bendat} {and} \bibinfo{person}{AG Piersol}.} \bibinfo{year}{1986}\natexlab{}.
\newblock \showarticletitle{Random data: Analysis and measurement procedures 2nd Edition A Wiley-Interscience Publication}.
\newblock \bibinfo{journal}{\emph{New York}} (\bibinfo{year}{1986}).
\newblock


\bibitem[Castera and Borhade(2025)]%
        {castera2025fluid}
\bibfield{author}{\bibinfo{person}{Michelle~R. Castera} {and} \bibinfo{person}{M.~B. Borhade}.} \bibinfo{year}{2025}\natexlab{}.
\newblock \showarticletitle{Fluid Management}.
\newblock In \bibinfo{booktitle}{\emph{StatPearls [Internet]}}. \bibinfo{publisher}{StatPearls Publishing}, \bibinfo{address}{Treasure Island (FL)}.
\newblock
\urldef\tempurl%
\url{https://www.ncbi.nlm.nih.gov/books/NBK532305/?utm_source=chatgpt.com}
\showURL{%
\tempurl}
\newblock
\shownote{[Updated 2025 Apr 29]}.


\bibitem[{Charder Medical}(2024)]%
        {charder2024ms6001}
\bibfield{author}{\bibinfo{person}{{Charder Medical}}.} \bibinfo{year}{2024}\natexlab{}.
\newblock \bibinfo{title}{MS6001 Digital Bed Scale}.
\newblock
\urldef\tempurl%
\url{https://www.chardermedical.com/bed-weighing-scales/MS6001.html}
\showURL{%
\tempurl}


\bibitem[Chelly et~al\mbox{.}(2024)]%
        {chelly2024trainable}
\bibfield{author}{\bibinfo{person}{Irit Chelly}, \bibinfo{person}{Shahaf~E Finder}, \bibinfo{person}{Shira Ifergane}, {and} \bibinfo{person}{Oren Freifeld}.} \bibinfo{year}{2024}\natexlab{}.
\newblock \showarticletitle{Trainable highly-expressive activation functions}. In \bibinfo{booktitle}{\emph{European Conference on Computer Vision}}. Springer, \bibinfo{pages}{200--217}.
\newblock


\bibitem[Codling et~al\mbox{.}(2021)]%
        {codling2021masshog}
\bibfield{author}{\bibinfo{person}{Jesse~R Codling}, \bibinfo{person}{Amelie Bonde}, \bibinfo{person}{Yiwen Dong}, \bibinfo{person}{Siyi Cao}, \bibinfo{person}{Akkarit Sangpetch}, \bibinfo{person}{Orathai Sangpetch}, \bibinfo{person}{Hae~Young Noh}, {and} \bibinfo{person}{Pei Zhang}.} \bibinfo{year}{2021}\natexlab{}.
\newblock \showarticletitle{MassHog: Weight-sensitive occupant monitoring for pig pens using actuated structural vibrations}. In \bibinfo{booktitle}{\emph{Adjunct Proceedings of the 2021 ACM International Joint Conference on Pervasive and Ubiquitous Computing and Proceedings of the 2021 ACM International Symposium on Wearable Computers}}. \bibinfo{pages}{600--605}.
\newblock


\bibitem[Davoodnia et~al\mbox{.}(2023)]%
        {davoodnia2023deep}
\bibfield{author}{\bibinfo{person}{Vandad Davoodnia}, \bibinfo{person}{Monet Slinowsky}, {and} \bibinfo{person}{Ali Etemad}.} \bibinfo{year}{2023}\natexlab{}.
\newblock \showarticletitle{Deep multitask learning for pervasive BMI estimation and identity recognition in smart beds}.
\newblock \bibinfo{journal}{\emph{Journal of Ambient Intelligence and Humanized Computing}} \bibinfo{volume}{14}, \bibinfo{number}{5} (\bibinfo{year}{2023}), \bibinfo{pages}{5463--5477}.
\newblock


\bibitem[Fr{\`y}ba(2013)]%
        {fryba2013vibration}
\bibfield{author}{\bibinfo{person}{Ladislav Fr{\`y}ba}.} \bibinfo{year}{2013}\natexlab{}.
\newblock \bibinfo{booktitle}{\emph{Vibration of solids and structures under moving loads}}. Vol.~\bibinfo{volume}{1}.
\newblock \bibinfo{publisher}{Springer science \& business media}.
\newblock


\bibitem[Greenwalt et~al\mbox{.}(2017)]%
        {greenwalt2017elimination}
\bibfield{author}{\bibinfo{person}{Mary Greenwalt}, \bibinfo{person}{David Griffen}, {and} \bibinfo{person}{Jim Wilkerson}.} \bibinfo{year}{2017}\natexlab{}.
\newblock \showarticletitle{Elimination of emergency department medication errors due to estimated weights}.
\newblock \bibinfo{journal}{\emph{BMJ Quality Improvement Reports}} \bibinfo{volume}{6}, \bibinfo{number}{1} (\bibinfo{year}{2017}).
\newblock


\bibitem[Harrington et~al\mbox{.}(2021)]%
        {harrington2021passive}
\bibfield{author}{\bibinfo{person}{Nicholas Harrington}, \bibinfo{person}{Quan~M Bui}, \bibinfo{person}{Zhe Wei}, \bibinfo{person}{Brandon Hernandez-Pacheco}, \bibinfo{person}{Pamela~N DeYoung}, \bibinfo{person}{Andrew Wassell}, \bibinfo{person}{Bayan Duwaik}, \bibinfo{person}{Akshay~S Desai}, \bibinfo{person}{Deepak~L Bhatt}, \bibinfo{person}{Parag Agnihotri}, {et~al\mbox{.}}} \bibinfo{year}{2021}\natexlab{}.
\newblock \showarticletitle{Passive longitudinal weight and cardiopulmonary monitoring in the home bed}.
\newblock \bibinfo{journal}{\emph{Scientific Reports}} \bibinfo{volume}{11}, \bibinfo{number}{1} (\bibinfo{year}{2021}), \bibinfo{pages}{24376}.
\newblock


\bibitem[He et~al\mbox{.}(2015)]%
        {he2015delving}
\bibfield{author}{\bibinfo{person}{Kaiming He}, \bibinfo{person}{Xiangyu Zhang}, \bibinfo{person}{Shaoqing Ren}, {and} \bibinfo{person}{Jian Sun}.} \bibinfo{year}{2015}\natexlab{}.
\newblock \showarticletitle{Delving deep into rectifiers: Surpassing human-level performance on imagenet classification}. In \bibinfo{booktitle}{\emph{Proceedings of the IEEE international conference on computer vision}}. \bibinfo{pages}{1026--1034}.
\newblock


\bibitem[{Health o meter Professional Scales}(2019)]%
        {healthometer2019accuracy}
\bibfield{author}{\bibinfo{person}{{Health o meter Professional Scales}}.} \bibinfo{year}{2019}\natexlab{}.
\newblock \bibinfo{title}{Accuracy Comparison between the Patient Transfer Scale, Typical ICU Bed, and Typical Stretcher}.
\newblock
\urldef\tempurl%
\url{https://www.homscales.com/innovations/patient-transfer-scale/}
\showURL{%
\tempurl}


\bibitem[Humar(2012)]%
        {humar2012dynamics}
\bibfield{author}{\bibinfo{person}{Jagmohan Humar}.} \bibinfo{year}{2012}\natexlab{}.
\newblock \bibinfo{booktitle}{\emph{Dynamics of structures}}.
\newblock \bibinfo{publisher}{CRC press}.
\newblock


\bibitem[Ishikawa et~al\mbox{.}(2022)]%
        {ishikawa2022long}
\bibfield{author}{\bibinfo{person}{Takahiro Ishikawa}, \bibinfo{person}{Ikuko Sakai}, \bibinfo{person}{Ayumi Amemiya}, \bibinfo{person}{Ryou Komatsu}, \bibinfo{person}{Shoko Sakuraba}, {and} \bibinfo{person}{Shiroh Isono}.} \bibinfo{year}{2022}\natexlab{}.
\newblock \showarticletitle{Long-term body weight change assessed by non-contact load cells under the bed in older people with and without eating assistance: a preliminary study}.
\newblock \bibinfo{journal}{\emph{Scientific Reports}} \bibinfo{volume}{12}, \bibinfo{number}{1} (\bibinfo{year}{2022}), \bibinfo{pages}{8107}.
\newblock


\bibitem[Kim and Hong(2020)]%
        {kim2020prediction}
\bibfield{author}{\bibinfo{person}{Tae-Hwan Kim} {and} \bibinfo{person}{Youn-Sik Hong}.} \bibinfo{year}{2020}\natexlab{}.
\newblock \showarticletitle{Prediction of body weight of a person lying on a smart mat in nonrestraint and unconsciousness conditions}.
\newblock \bibinfo{journal}{\emph{Sensors}} \bibinfo{volume}{20}, \bibinfo{number}{12} (\bibinfo{year}{2020}), \bibinfo{pages}{3485}.
\newblock


\bibitem[Krauss and Green(2006)]%
        {krauss2006procedural}
\bibfield{author}{\bibinfo{person}{Baruch Krauss} {and} \bibinfo{person}{Steven~M Green}.} \bibinfo{year}{2006}\natexlab{}.
\newblock \showarticletitle{Procedural sedation and analgesia in children}.
\newblock \bibinfo{journal}{\emph{The Lancet}} \bibinfo{volume}{367}, \bibinfo{number}{9512} (\bibinfo{year}{2006}), \bibinfo{pages}{766--780}.
\newblock


\bibitem[Lubitz et~al\mbox{.}(1988)]%
        {lubitz1988rapid}
\bibfield{author}{\bibinfo{person}{Deborah~S Lubitz}, \bibinfo{person}{James~S Seidel}, \bibinfo{person}{Leon Chameides}, \bibinfo{person}{Robert~C Luten}, \bibinfo{person}{Arno~L Zaritsky}, {and} \bibinfo{person}{Frederick~W Campbell}.} \bibinfo{year}{1988}\natexlab{}.
\newblock \showarticletitle{A rapid method for estimating weight and resuscitation drug dosages from length in the pediatric age group}.
\newblock \bibinfo{journal}{\emph{Annals of emergency medicine}} \bibinfo{volume}{17}, \bibinfo{number}{6} (\bibinfo{year}{1988}), \bibinfo{pages}{576--581}.
\newblock


\bibitem[Ma et~al\mbox{.}(2021)]%
        {ma2021activate}
\bibfield{author}{\bibinfo{person}{Ningning Ma}, \bibinfo{person}{Xiangyu Zhang}, \bibinfo{person}{Ming Liu}, {and} \bibinfo{person}{Jian Sun}.} \bibinfo{year}{2021}\natexlab{}.
\newblock \showarticletitle{Activate or not: Learning customized activation}. In \bibinfo{booktitle}{\emph{Proceedings of the IEEE/CVF Conference on Computer Vision and Pattern Recognition}}. \bibinfo{pages}{8032--8042}.
\newblock


\bibitem[{Massachusetts Department of Public Health}(2025)]%
        {massEMS2025}
\bibfield{author}{\bibinfo{person}{{Massachusetts Department of Public Health}}.} \bibinfo{year}{2025}\natexlab{}.
\newblock \bibinfo{title}{Emergency Medical Services Pre-Hospital Statewide Treatment Protocols}.
\newblock \bibinfo{howpublished}{\url{https://www.mass.gov/doc/emergency-medical-services-statewide-treatment-protocols-version-20251-effective-june-16-2025-0/download}}.
\newblock
\newblock
\shownote{Version 2025.1, Effective June 16, 2025}.


\bibitem[Mehta(2020)]%
        {mehta2020accuracy}
\bibfield{author}{\bibinfo{person}{Renuka Mehta}.} \bibinfo{year}{2020}\natexlab{}.
\newblock \showarticletitle{The accuracy of the Broselow Tape in overweight and obese patients}.
\newblock \bibinfo{journal}{\emph{Pediatrics}} \bibinfo{volume}{146}, \bibinfo{number}{1\_MeetingAbstract} (\bibinfo{year}{2020}), \bibinfo{pages}{158--159}.
\newblock


\bibitem[Molina et~al\mbox{.}(2019)]%
        {molina2019pad}
\bibfield{author}{\bibinfo{person}{Alejandro Molina}, \bibinfo{person}{Patrick Schramowski}, {and} \bibinfo{person}{Kristian Kersting}.} \bibinfo{year}{2019}\natexlab{}.
\newblock \showarticletitle{Pad$\backslash$'e activation units: End-to-end learning of flexible activation functions in deep networks}.
\newblock \bibinfo{journal}{\emph{arXiv preprint arXiv:1907.06732}} (\bibinfo{year}{2019}).
\newblock


\bibitem[Ramachandran et~al\mbox{.}(2017)]%
        {ramachandran2017swish}
\bibfield{author}{\bibinfo{person}{Prajit Ramachandran}, \bibinfo{person}{Barret Zoph}, {and} \bibinfo{person}{Quoc~V Le}.} \bibinfo{year}{2017}\natexlab{}.
\newblock \showarticletitle{Swish: a self-gated activation function}.
\newblock \bibinfo{journal}{\emph{arXiv preprint arXiv:1710.05941}} \bibinfo{volume}{7}, \bibinfo{number}{1} (\bibinfo{year}{2017}), \bibinfo{pages}{5}.
\newblock


\bibitem[Shafi et~al\mbox{.}(2022)]%
        {shafi2022design}
\bibfield{author}{\bibinfo{person}{Imran Shafi}, \bibinfo{person}{Muhammad~Siddique Farooq}, \bibinfo{person}{Isabel De~La Torre~D{\'\i}ez}, \bibinfo{person}{Jose Bre{\~n}osa}, \bibinfo{person}{Julio C{\'e}sar~Mart{\'\i}nez Espinosa}, {and} \bibinfo{person}{Imran Ashraf}.} \bibinfo{year}{2022}\natexlab{}.
\newblock \showarticletitle{Design and development of smart weight measurement, lateral turning and transfer bedding for unconscious patients in pandemics}. In \bibinfo{booktitle}{\emph{Healthcare}}, Vol.~\bibinfo{volume}{10}. MDPI, \bibinfo{pages}{2174}.
\newblock


\bibitem[Sundararajan and Najmi(2020)]%
        {sundararajan2020many}
\bibfield{author}{\bibinfo{person}{Mukund Sundararajan} {and} \bibinfo{person}{Amir Najmi}.} \bibinfo{year}{2020}\natexlab{}.
\newblock \showarticletitle{The many Shapley values for model explanation}. In \bibinfo{booktitle}{\emph{International conference on machine learning}}. PMLR, \bibinfo{pages}{9269--9278}.
\newblock


\bibitem[{Suno AI}(2023)]%
        {sunoai2023}
\bibfield{author}{\bibinfo{person}{{Suno AI}}.} \bibinfo{year}{2023}\natexlab{}.
\newblock \bibinfo{title}{Suno: AI Music Generator}.
\newblock
\urldef\tempurl%
\url{https://www.suno.ai/}
\showURL{%
\tempurl}
\newblock
\shownote{Accessed: 2025-08-31}.


\bibitem[S{\"u}tfeld et~al\mbox{.}(2020)]%
        {sutfeld2020adaptive}
\bibfield{author}{\bibinfo{person}{Leon~Ren{\'e} S{\"u}tfeld}, \bibinfo{person}{Flemming Brieger}, \bibinfo{person}{Holger Finger}, \bibinfo{person}{Sonja F{\"u}llhase}, {and} \bibinfo{person}{Gordon Pipa}.} \bibinfo{year}{2020}\natexlab{}.
\newblock \showarticletitle{Adaptive blending units: Trainable activation functions for deep neural networks}. In \bibinfo{booktitle}{\emph{Science and Information Conference}}. Springer, \bibinfo{pages}{37--50}.
\newblock


\bibitem[Trottier et~al\mbox{.}(2017)]%
        {trottier2017parametric}
\bibfield{author}{\bibinfo{person}{Ludovic Trottier}, \bibinfo{person}{Philippe Giguere}, {and} \bibinfo{person}{Brahim Chaib-Draa}.} \bibinfo{year}{2017}\natexlab{}.
\newblock \showarticletitle{Parametric exponential linear unit for deep convolutional neural networks}. In \bibinfo{booktitle}{\emph{2017 16th IEEE international conference on machine learning and applications (ICMLA)}}. IEEE, \bibinfo{pages}{207--214}.
\newblock


\bibitem[Vieira and Kumar(2009)]%
        {vieira2009safety}
\bibfield{author}{\bibinfo{person}{ER Vieira} {and} \bibinfo{person}{S Kumar}.} \bibinfo{year}{2009}\natexlab{}.
\newblock \showarticletitle{Safety analysis of patient transfers and handling tasks}.
\newblock \bibinfo{journal}{\emph{BMJ Quality \& Safety}} \bibinfo{volume}{18}, \bibinfo{number}{5} (\bibinfo{year}{2009}), \bibinfo{pages}{380--384}.
\newblock


\bibitem[Vinter et~al\mbox{.}(2023)]%
        {vinter2023electrical}
\bibfield{author}{\bibinfo{person}{Nicoline Vinter}, \bibinfo{person}{Mads Z.~B. Holst-Hansen}, \bibinfo{person}{S{\o}ren~P. Johnsen}, \bibinfo{person}{Gregory Y.~H. Lip}, \bibinfo{person}{Lars Frost}, {and} \bibinfo{person}{Ludovic Trinquart}.} \bibinfo{year}{2023}\natexlab{}.
\newblock \showarticletitle{Electrical energy by electrode placement for cardioversion of atrial fibrillation: a systematic review and meta-analysis}.
\newblock \bibinfo{journal}{\emph{Open Heart}} \bibinfo{volume}{10}, \bibinfo{number}{2} (\bibinfo{date}{Nov.} \bibinfo{year}{2023}), \bibinfo{pages}{e002456}.
\newblock
\href{https://doi.org/10.1136/openhrt-2023-002456}{doi:\nolinkurl{10.1136/openhrt-2023-002456}}


\bibitem[{VPG Force Sensors}(2024)]%
        {vpgforcesensors2024hospital}
\bibfield{author}{\bibinfo{person}{{VPG Force Sensors}}.} \bibinfo{year}{2024}\natexlab{}.
\newblock \bibinfo{title}{Hospital Beds with Integrated Load Cells for Patient Weighing}.
\newblock \bibinfo{howpublished}{VPG Force Sensors Technical Blog}.
\newblock
\urldef\tempurl%
\url{https://blog.vpgforcesensors.com/hospital-beds-with-integrated-load-cells-for-patient-weighing/}
\showURL{%
\tempurl}
\newblock
\shownote{Accessed August 7, 2025}.


\bibitem[Wells et~al\mbox{.}(2013)]%
        {wells2013pawper}
\bibfield{author}{\bibinfo{person}{Mike Wells}, \bibinfo{person}{Ashraf Coovadia}, \bibinfo{person}{Efraim Kramer}, {and} \bibinfo{person}{Lara Goldstein}.} \bibinfo{year}{2013}\natexlab{}.
\newblock \showarticletitle{The PAWPER tape: a new concept tape-based device that increases the accuracy of weight estimation in children through the inclusion of a modifier based on body habitus}.
\newblock \bibinfo{journal}{\emph{Resuscitation}} \bibinfo{volume}{84}, \bibinfo{number}{2} (\bibinfo{year}{2013}), \bibinfo{pages}{227--232}.
\newblock


\bibitem[Wells et~al\mbox{.}(2017)]%
        {wells2017accuracy}
\bibfield{author}{\bibinfo{person}{Mike Wells}, \bibinfo{person}{Lara~Nicole Goldstein}, {and} \bibinfo{person}{Alison Bentley}.} \bibinfo{year}{2017}\natexlab{}.
\newblock \showarticletitle{The accuracy of emergency weight estimation systems in children—a systematic review and meta-analysis}.
\newblock \bibinfo{journal}{\emph{International Journal of Emergency Medicine}}  \bibinfo{volume}{10} (\bibinfo{year}{2017}), \bibinfo{pages}{1--43}.
\newblock


\bibitem[Wells and Yende(2023)]%
        {wells2023there}
\bibfield{author}{\bibinfo{person}{Mike Wells} {and} \bibinfo{person}{Penelope Yende}.} \bibinfo{year}{2023}\natexlab{}.
\newblock \showarticletitle{Is there evidence that length-based tapes with precalculated drug doses increase the accuracy of drug dose calculations in children? A systematic review}.
\newblock \bibinfo{journal}{\emph{Clinical and Experimental Emergency Medicine}} \bibinfo{volume}{11}, \bibinfo{number}{2} (\bibinfo{year}{2023}), \bibinfo{pages}{145}.
\newblock


\bibitem[Wu et~al\mbox{.}(2023)]%
        {wu2023massnet}
\bibfield{author}{\bibinfo{person}{Ziyu Wu}, \bibinfo{person}{Quan Wan}, \bibinfo{person}{Mingjie Zhao}, \bibinfo{person}{Yi Ke}, \bibinfo{person}{Yiran Fang}, \bibinfo{person}{Zhen Liang}, \bibinfo{person}{Fangting Xie}, {and} \bibinfo{person}{Jingyuan Cheng}.} \bibinfo{year}{2023}\natexlab{}.
\newblock \showarticletitle{Massnet: A deep learning approach for body weight extraction from a single pressure image}. In \bibinfo{booktitle}{\emph{2023 IEEE International Conference on Pervasive Computing and Communications (PerCom)}}. IEEE, \bibinfo{pages}{180--189}.
\newblock


\bibitem[Zhang et~al\mbox{.}(2025)]%
        {zhang2025wevibe}
\bibfield{author}{\bibinfo{person}{Jiale Zhang}, \bibinfo{person}{Yuyan Wu}, \bibinfo{person}{Jesse~R Codling}, \bibinfo{person}{Yen~Cheng Chang}, \bibinfo{person}{Julia Gersey}, \bibinfo{person}{Pei Zhang}, \bibinfo{person}{Hae~Young Noh}, {and} \bibinfo{person}{Yiwen Dong}.} \bibinfo{year}{2025}\natexlab{}.
\newblock \showarticletitle{WeVibe: Weight Change Estimation Through Audio-Induced Shelf Vibrations In Autonomous Stores}.
\newblock \bibinfo{journal}{\emph{arXiv preprint arXiv:2502.12093}} (\bibinfo{year}{2025}).
\newblock


\bibitem[Zhang et~al\mbox{.}(2020)]%
        {zhang2020vibroscale}
\bibfield{author}{\bibinfo{person}{Shibo Zhang}, \bibinfo{person}{Qiuyang Xu}, \bibinfo{person}{Sougata Sen}, {and} \bibinfo{person}{Nabil Alshurafa}.} \bibinfo{year}{2020}\natexlab{}.
\newblock \showarticletitle{VibroScale: Turning your smartphone into a weighing scale}. In \bibinfo{booktitle}{\emph{Adjunct Proceedings of the 2020 ACM International Joint Conference on Pervasive and Ubiquitous Computing and Proceedings of the 2020 ACM International Symposium on Wearable Computers}}. \bibinfo{pages}{176--179}.
\newblock


\end{thebibliography}







\end{document}